\documentclass{article}
\pdfoutput=1
\usepackage{url}
\usepackage{amssymb, amsfonts, amsmath}
\usepackage{graphicx, natbib}
\usepackage{url}
\usepackage{amssymb, amsfonts, amsmath}
\usepackage{graphicx, natbib}

\newcommand{\BIT}{\begin{itemize}}
\newcommand{\EIT}{\end{itemize}}
\newcommand{\BNUM}{\begin{enumerate}}
\newcommand{\ENUM}{\end{enumerate}}

\newcommand\mbb[1]{\mathbb{#1}}

\def\reals{\mathbb{R}} 
\def\simplex{\mathcal{S}} 
\renewcommand{\exp}[1]{\operatorname{exp}\left(#1\right)} 
\def\indic#1{\mbb{I}\left({#1}\right)} 

\providecommand{\asinh}{\mathop\mathrm{asinh}}

\def\E{\mathbb{E}} 
\def\Earg#1{\E\left[{#1}\right]}

\def\P{\mathbb{P}} 
\def\Parg#1{\P\left({#1}\right)}

\def\Gsn{\mathcal{N}}

\def\Mult{\textnormal{Mult}}

\def\Gam{\textnormal{Gam}}

\def\Dir{\textnormal{Dir}}

\def\Poi{\textnormal{Poi}}

\begin{document}

\title{Latent Variable Modeling for the Microbiome}

\author{
  Kris Sankaran and Susan P. Holmes \\
  \textit{Stanford University}
}

\markboth
{Sankaran and Holmes}
{Latent Variable Modeling for the Microbiome}

\maketitle


\begin{abstract}
  {
    The human microbiome is a complex ecological system, and describing its
    structure and function under different environmental conditions is important
    from both basic scientific and medical perspectives. Viewed through a
    biostatistical lens, many microbiome analysis goals can be formulated as
    latent variable modeling problems. However, although probabilistic latent
    variable models are a cornerstone of modern unsupervised learning, they are
    rarely applied in the context of microbiome data analysis, in spite of the
    evolutionary, temporal, and count structure that could be directly
    incorporated through such models. We explore the application of
    probabilistic latent variable models to microbiome data, with a focus on
    Latent Dirichlet Allocation, Nonnegative Matrix Factorization, and Dynamic
    Unigram models. To develop guidelines for when different methods are
    appropriate, we perform a simulation study. We further illustrate and
    compare these techniques using the data of \cite{dethlefsen2011incomplete},
    a study on the effects of antibiotics on bacterial community composition.
    Code and data for all simulations and case studies are available publicly.}
\end{abstract}

\section{Introduction}

Microbiome studies attempt to characterize variation in bacterial abundance
profiles across different experimental conditions \citep{human2012structure,
  gilbert2014earth}. For example, a study may attempt to describe differences in
bacterial communities between diseased and healthy states or after deliberately
induced perturbations \citep{dethlefsen2011incomplete, fukuyama2017multidomain}.

In the process, two complementary difficulties arise. First, the data are often
high-dimensional, measured over several hundreds or thousands types of bacteria.
Studying patterns at the level of particular bacteria is typically
uninformative. Second, it can be important to study bacterial abundances in the
context of existing biological knowledge. For example, it is scientifically
meaningful when a collection of bacteria that are known to be evolutionarily
related have either similar or opposed abundance profiles. These challenges
motivate methodological work in the microbiome, including
phylogenetically-informed techniques for dimensionality-reduction and
high-dimensional regression.

Towards phylogenetically-informed dimensionality reduction,
\cite{chen2013structure}, who applied a structured regularization penalty in
sparse CCA to incorporate phylogenetic information, encouraging scores for
similar species to be placed close to one another. A similar effect can be
obtained by placing an appropriate prior in a Bayesian model, an approach
explored in \cite{fukuyama2017adaptive}. By varying the strength of the prior,
it is possible to encourage different degrees of smoothness with respect to
phylogeny, and Empirical Bayes estimation allows for a certain type of
adaptivity.

In addition to dimensionality reduction, high-dimensional classification is
popular in the microbiome literature. Often interesting species can be
identified by searching for important features in models that predict sample
characteristics -- treatment vs. control, for example -- from species
abundances. \cite{segata2011metagenomic} provided an approach for accounting for
phylogenetic structure in this problem by initially prescreening according to
independent biological interest. Alternatively, \cite{chen2013variable} applied
a Dirichlet Multinomial regression model to study the relationship between
bacterial counts and sample characteristics in a fully generative fashion, and
then added an $\ell^{1}$-penalty to induce sparsity and facilitate
interpretability. \cite{shafiei2015biomico} designed a different generative
supervised model, based on ideas from mixed-membership modeling
\citep{blei2003latent}. The development of structured modeling techniques
tailored to the microbiome remains an active area of current research.

\section{Methods}

We review a few of the statistical modeling techniques that are the focus of
this work. Many of these techniques have been borrowed from text analysis,
thinking of the bacterial counts matrix as a biological analog of the
document-term matrix. The idea of transferring these techniques to the
microbiome is not new, though its appropriateness and usefulness has only been
explored in relatively limited settings \citep{schloss2007last,
  holmes2012dirichlet, chen2012estimating, chen2013variable,
  shafiei2015biomico, jiang2017microbiome}.

\subsection{Dirichlet Multinomial Mixture Model}

While the multinomial distribution is the fundamental probability mechanism for
sampling counts, multinomial models are only appropriate for relatively
homogeneous data sets, where categories are nearly independent. An extension,
Dirichlet multinomial mixture modeling, allows for count modeling in the
presence of increased heterogeneity, and its relevance to the text and
microbiome settings were key contributions of \cite{nigam2000text} and
\cite{holmes2012dirichlet}, respectively. In this section, we adopt text
analysis terminology, where the count matrices of interest are counts of terms
across documents. In Section \ref{sec:microbiome_vs_text_analysis} we clarify
the connection to microbiome analysis.

Suppose there are $D$ documents across $V$ terms, and that these documents are
assumed to belong to one of $K$ underlying topics, where a topic is defined as a
distribution over words. The Dirichlet multinomial mixture model draws each
topic from a distribution over probabilities. Then, for each document, a topic
is chosen by flipping a $K$-sided coin with probability $\theta_{k}$ of coming
up on side $k$. Conditional on the selected topic, all words are drawn
independently from the probabilites associated with the selected topic.

More formally, represent the topic for the $d^{th}$ document by $z_{d} \in \{1,
\dots, K\}$ and the term in the $n_{th}$ word of this document by $w_{dn}$.
Suppose the $k^{th}$ topic places weight $\beta_{vk}$ on the $v^{th}$ term, so
that $\beta_{k} \in \simplex^{V - 1}$. Suppose there are $N_{d}$ words in the
$d^{th}$ document. Then, the generative mechanism for the observed data set is
\begin{align*}
  w_{dn} \vert \left(\beta_{k}\right)_{k = 1}^{K}, z_{d} &\overset{iid}{\sim}\Mult\left(1, \beta_{z_{d}}\right) \text{ for } d = 1, \dots, D \text{ and } n = 1, \dots, N_{d} \\
  z_{d} \vert \theta &\overset{iid}{\sim} \Mult\left(1, \theta\right) \text{ for } d = 1, \dots, D \\
  \beta_{k} &\overset{iid}{\sim} \Dir\left(\gamma\right) \text{ for } k = 1, \dots, K.
\end{align*}
Equivalently, we could write $w_{d} \vert \left(\beta_{k}\right)_{k = 1}^{K},
z_{d} \sim \Mult\left(N_{d}, \beta_{z_{d}}\right)$, though the form above makes
comparison with LDA more straightforward. Further, while we treat $\theta$, and
$\gamma$ as fixed parameters, it is possible to place priors on them as well.
Geometrically, this model identifies each document with one of the $\beta_{k}$ on
the $V$-dimensional simplex.

In practice, interpretation revolves around the posterior topic memberships,
$\Parg{z_{d} = k \vert \left(w_{dn}\right)}$, and probabilities,
$\Parg{\beta_{kv}\vert \left(w_{dn}\right)}$. While these
estimates can be useful in guiding scientific analysis \citep{nigam2000text,
  holmes2012dirichlet}, the assumption that each document belongs completely to
one topic is sometimes unnecessarily restrictive. For example, in learning a
Dirichlet multinomial mixture model on a collection of newspaper articles, we
may recover separate topics related to science and personal health, but there
would be no way to express the mixture of topics in an article about health
research.

\subsection{Latent Dirichlet Allocation}

Latent Dirichlet Allocation (LDA) is a generalization of Dirichlet multinomial
mixture modeling where documents are allowed to have fractional membership
across a set of topics \citep{blei2003latent}. This addresses the key limitation
of Dirichlet multinomial mixture modeling, and one goal of the case study in
Section \ref{sec:data_analysis} is to demonstrate the usefulness of this
additional flexibility in microbiome analysis.

We use the same notation as before, but instead of fixing a global $\theta$
parameter, let $\theta_{d} \in \simplex^{K - 1}$ represent the $d^{th}$
document's mixture over the $K$ underlying topics. Represent the term in the
$n_{th}$ word of this document by $w_{dn}$, and the associated topic by
$z_{dn}$. Suppose the $k^{th}$ topic places weight $\beta_{vk}$ on the $v^{th}$
term, so that $\beta_{k} \in \simplex^{V - 1}$. Suppose there are $N_{d}$ words
in the $d^{th}$ document. Then, the generative mechanism for the observed data
set is
\begin{align*}
w_{dn} \vert \left(\beta_{k}\right)_{k = 1}^{K}, z_{dn} &\overset{iid}{\sim} \Mult\left(1, \beta_{z_{dn}}\right) \text{ for } d = 1, \dots, D  \text{ and } n = 1, \dots,  N_{d} \\
z_{dn} \vert \theta_{d} &\overset{iid}{\sim} \Mult\left(1, \theta_{d}\right) \text{ for } d = 1, \dots, D \text{ and } n = 1, \dots, N_{d}\\
\theta_{d} &\overset{iid}{\sim} \Dir\left(\alpha\right) \text{ for } d = 1, \dots, D \\
\beta_{k} &\overset{iid}{\sim} \Dir\left(\gamma\right) \text{ for } k = 1, \dots, K.
\end{align*}
In microbiome applications, we will find a formulation that marginalizes over
the $z_{dn}$ more convenient. Setting $x_{dv} = \sum_{n = 1}^{N_{d}}
\indic{w_{dn} = v}$, we can write the marginal distribution as
\begin{align*}
x_{d\cdot} \vert \left(\beta_{k}\right)_{1}^{K} &\overset{iid}{\sim} \Mult\left(N_{d}, B\theta_{d}\right) \text{ for } d = 1, \dots, D\\
\theta_{d} &\overset{iid}{\sim} \Dir\left(\alpha\right) \text{ for } d = 1, \dots, D \\
\beta_{k} &\overset{iid}{\sim} \Dir\left(\gamma\right) \text{ for }k = 1, \dots, K,
\end{align*}
where we have introduced the $V \times K$ matrix concatenating all topics
column-wise, $B = \begin{pmatrix}\beta_{1}, \dots, \beta_{K}\end{pmatrix}$.
Geometrically, LDA identifies samples with points in the convex hull of $K$
topics $\left(\beta_{k}\right)$ on the $V$-dimensional simplex, rather than the
individual corners, as in the Dirichlet multinomial mixture model.

\subsection{Dynamic Unigram model}

Upon examining this geometric interpretation, we might consider in some
situations a model that identifies samples with a continuous curve on this
$V$-dimensional simplex. This reasoning leads naturally to the Dynamic
Unigram model \citep{blei2006dynamic}. The underlying curve reflects the gradual evolution
of probabilities over time, and is implemented by passing a random walk through
a multilogit link. That is, the Dynamic Unigram model posits the generative
model
\begin{align*}
x_{d\cdot} \vert \mu_{t\left(d\right)}  &\overset{iid}{\sim} \Mult\left(N_{d}, S\left(\mu_{t\left(d\right)}\right)\right) \text{ for } d = 1, \dots, D\\
\mu_{t} \vert \mu_{t - 1} &\overset{iid}{\sim} \Gsn\left(\mu_{t - 1}, \sigma^{2}I_{V}\right) \text{ for } t = 1, \dots, T \\
\mu_{0} &\overset{iid}{\sim} \Gsn\left(0, \sigma^{2}I_{V}\right),
\end{align*}
where $S$ is the multilogit link
\begin{align*}
\left[S\left(\mu\right)\right]_{v} = \frac{\exp{\mu_{v}}}{\sum_{v^{\prime}} \exp{\mu_{v^{\prime}}}},
\end{align*}
and $t\left(d\right)$ maps document $d$ to the time it was sampled. The
$\left(\mu_{t\left(\right)}\right)$ define a Gaussian random walk in
$\reals^{V}$ with step-size $\sigma$, and $S$ transforms the walk into a
sequence of probability distributions. In our experiments, we place a vague
inverse-gamma prior on $\sigma^{2}$, since this hyperparameter is rarely known
in practice.

\subsection{Nonnegative Matrix Factorization}
\label{sec:nmf}

In LDA, count matrices are modeled by sampling from multinomials with total
counts coming from the word count of each document and probabilities
coming from the rows of $\Theta B^{T}$ where $\Theta = \begin{pmatrix}\theta_{1}
  \\ \vdots \\ \theta_{D} \end{pmatrix}$ and $B = \begin{pmatrix} \beta_{1}
  \dots \beta_{K} \end{pmatrix}$ are $D \times K$ and $V \times K$ matrices
representing document and topic distributions, respectively, and where each
$\theta_{d} \in \simplex^{K - 1}$ and $\beta_{k} \in \simplex^{V - 1}$.

Alternatively, it is possible to model the nonnegative matrix $X$ by the product
of low rank matrices, $X \approx \Theta B^{T}$, where now the only constraints
on $\Theta$ and $B$ are that $\Theta \in \reals_{+}^{D \times K}$ and $B \in
\reals_{+}^{V \times K}$. This is the point of departure for a variety of algorithms
in the Nonnegative Matrix Factorization (NMF) literature
\citep{wang2013nonnegative, berry2007algorithms, lee2001algorithms}

We focus on the Gamma-Poisson factorization model (GaP)
\citep{kucukelbir2015automatic, carpenter2016stan, zhou2015negative,
  canny2004gap} which posits the hierarchical model
\begin{align*}
X &\sim \Poi\left(\Theta B^{T}\right) \\
\Theta &\sim \Gam\left(a_{0} 1_{D \times K}, b_{0} 1_{D \times K}\right) \\
B &\sim \Gam\left(c_{0} 1_{V \times K}, d_{0} 1_{V \times K} \right),
\end{align*}
where our notation means that each entry in these matrices is sampled
independently, with parameters given by the corresponding entry in the parameter
matrix. As a consequence of the representation of the negative binomial
distribution as a Gamma mixture of Poissons, this is a natural model of
overdispersed counts, which arise frequently in genomic and microbiome settings
\citep{love2014moderated, mcmurdie2014waste}. In practice, the hyperparameters
$a_{0}, b_{0}, c_{0}$, and $d_{0}$ are unknown. In all of our experiments, we
optimize over them, though it would also be possible to add a level to the
hierarchy and place vague nonnegative priors on these parameters.

In our simulation studies, we also consider a slight variant of this model,
similar to the proposal of \citep{romero2014composition}, which independently sends
entries of $X$ to zero with probability $p_{0}$. In our experiments, we assume
that this zero-inflation probability is known. This procedure is denoted by
Z-NMF.

\subsection{Posterior Predictive Checks}
\label{sec:ppc_overview}

Model assessment is important for qualifying interpretations, and can further
guide refinements in subsequent analyses. Indeed, part of the appeal of
probabilistic modeling is the ease with which models can be adapted to better
describe the data of interest. We briefly review model assessment via posterior
predictive checks \citep{rubin1984bayesianly, gelman2013philosophy}, as they are
applied in Section \ref{sec:antibiotics_ppc}. In this approach, some statistics
$T_{k}\left(x\right)$ of the data are defined which, in some sense,
``characterize'' the data. If the data $x^{\ast}$ simulated from the fitted
model have statistics $T_{k}\left(x^{\ast}\right)$ with values similar to those
in the observed data $T_{k}\left(x\right)$, then we have evidence that the
proposed model approximates the data well, at least in the sense defined by
$T_{k}$.

More precisely, simulate data $x_{1}^{\ast}, \dots, x_{S}^{\ast}$ from the
posterior predictive probability distribution $p\left(x^{\ast}\vert x\right)
\approx \int p\left(x^{\ast} \vert \theta\right) \hat{p}\left(\theta \vert x
\right)d\theta$, where $x$ is the original data and $\hat{p}\left(\theta \vert
x\right)$ is an estimate of the posterior. For each of these simulated data
sets, the characterizing statistics $T_{k}\left(x_{s}\right)$ are computed.
Graphically comparing the $T_{k}\left(x\right)$ calculated on the true data with
the histogram of model-fit simulated $T_{k}\left(x_{s}^{\ast}\right)$ suggests
ways in which the posited model fits -- the case where the observed
$T_{k}\left(x\right)$ lie in the bulk of the $T_{k}\left(x^{\ast}_{s}\right)$ --
or fails to fit -- the case where $T_{k}\left(x\right)$ lie far from the bulk of
$T_{k}\left(x^{\ast}_{s}\right)$ -- the data well.

For example, it is common to set $T_{k}\left(x\right) = \overline{x}_d$ or
$\frac{1}{n}\sum_{i} \left(x_{id} - \overline{x}_d\right)^{2}$ to see whether
simulated samples approximately match the moments of the $d^{th}$ dimension in
the observed data. Alternatively, histograms of raw data or raw data subsetted
to certain groups can guide evaluation. This corresponds to setting a
multidimensional $T_{k}\left(x\right)$. For example, $T_k\left(x_{s}^\ast\right)
= \left(n_{s1}, \dots, n_{sB}\right)$ could count the number of observations in
the $s^{th}$ simulated data set falling into histogram bins $b = 1, \dots, B$.

\subsection{Microbiome vs. Text Analysis}
\label{sec:microbiome_vs_text_analysis}

One of the primary contributions of our work is to develop the observation that
methods popular in text analysis can be adapted to the microbiome setting in a
way that produces useful summaries. Before applying these methods, we develop
the analogy between these text and microbiome analysis and also draw attention
to points where the parallels break down.

In the abstract, it becomes clear that the semantic differences between the
units of statistical analysis are often superficial. For example, we can map
between the most common field-specific terms as follows,

\begin{itemize}
  \item \textbf{Document} $\iff$ \textbf{Biological Sample}: The basic sampling
    units, over which conclusions are generalized, are documents (text analysis)
    and biological samples (microbiome analysis). It is of interest to highlight
    similarities and differences across these units, often through some
    variation on clustering or dimensionality reduction.
  \item \textbf{Term} $\iff$ \textbf{Bacterial species}: The fundamental
    features with which to describe samples are the counts of terms (text
    analysis) and bacterial species (microbiome analysis). More formally, by
    bacterial species, we mean Amplicon Sequence Variants
    \citep{callahan2017exact}, though we avoid this terminology for simplicity
    of exposition.
  \item \textbf{Topic} $\iff$ \textbf{Community}: For interpretation, it is
    common to imagine ``prototypical'' units which can be used as a point of
    reference for observed samples. In text analysis, these are called topics --
    for example, ``business'' or ``politics'' articles have their own specific
    vocabularies. On the other hand, in microbiome analysis, these are called
    ``communities'' -- different communities have different bacterial
    signatures.
  \item \textbf{Word} $\iff$ \textbf{Sequencing Read}: A ``word'' in text analysis refers
    to a single instance of a term in a document, not its total count. The
    analog in microbiome analysis is an individual read that has been mapped to
    a unique sequence variant, though this is rarely an object of intrinsic interest.
  \item \textbf{Corpus} $\iff$ \textbf{Environment}: Sometimes a grouping
    structure is known apriori among sampling units. In this case, it can be
    informative to describe whether topics are or are not shared across these
    groups \citep{teh2004sharing}. In the text analysis literature, a known group
    of documents -- for example, all articles coming from one newspaper -- is
    called a corpus. In the microbiome literature, the associated concept is the
    environment -- for example, skin, ocean, or soil -- from which a sample was
    obtained.
\end{itemize}

Now that we have established the semantic connections between text and
microbiome analysis, we compare the types of data and analysis goals that are
typical within the respective fields. In both, a central element of study is
the sample-by-feature (either document-by-term or biological sample-by-bacteria)
matrix. Besides count structure, the most striking similarity between these data
matrices is sparsity: most entries are zero. Further, observed counts tend to be
highly-skewed -- some terms are far more common than others, and in the same
way, some microbes are much more abundant than others. Finally, in both fields,
contextual information beyond the raw sample-by-feature matrix is typically
available. For example, timestamps are often available in both domains,
$n$-grams have a natural analog in terms of small subnetworks of co-occurring
bacteria, and phylogenetic similarity between species parallels apriori known
linguistic characteristics of terms.

Nonetheless, in practice, the structure of these data can be quite different.
First, text data can be on a much larger scale. For example, the Wikipedia
corpus studied by \cite{hoffman2013stochastic} includes 3.8 million articles. In
contrast, even large microbiome datasets, like the one studied in
\citep{gilbert2014earth}, typically only have on the order of tens of thousands
of samples. Similarly, the total number of terms in such large-scale text
analysis problems can be substantially larger than the number of bacterial
species under consideration.

On the other hand, in a different sense, microbiome studies are larger in scale
-- there tend to be tens of thousands of reads per sample in microbiome studies,
but only hundreds to thousands of words within any article. This means that
techniques that rely on the representation of documents as sequences of
individual words, rather than vectors of word counts, require too much memory to
be practical. This makes many useful text analysis techniques -- for example,
some methods for model inference \citep{griffiths2004finding} and evaluation
\citep{wallach2009evaluation} -- out of reach for standard microbiome problems.
This does, however, suggest potential opportunities for future research.

Lastly, we compare the prevailing analytical goals within text and microbiome
analysis. In both fields, data reduction can be informative for developing
models of system behavior. However, an essential difference is that even
unsupervised text analysis techniques are often embedded within automatic
systems, for text classification \citep{blei2003latent} or information retrieval
\citep{krestel2009latent} for example, which do not require the intervention of
a scientific investigator. In contrast, in microbiome studies, researchers often
have control over specific experimental design structure, and collect and
analyze data on a per-study basis. In this setting, success is defined somewhat
amorphously as an ability to describe the structure and function underlying a
biological system of interest.

\section{Simulation Study}

It can be liberating to have easy access to such a variety of modeling
strategies for any given microbiome analysis problem. However, with this
increased flexibility comes the difficulty of determining when to use which
methods. To build some intuition about estimation accuracy across combinations
of data settings and model types, we conduct a series of simulation studies.
These are meant to complement the model-checking that should follow parametric
analysis -- since we know the truth in simulations, it is easier to develop more
unambiguous guidelines.

More specifically, our plan is to divide our experiments into simulations
generating data from the true LDA, unigram, and NMF / Z-NMF models. In each, we
vary the sample size and dimension, performing model estimation using either
Markov Chain Monte Carlo (MCMC) sampling, Variational Bayes (VB), or a
bootstrap. The only misspecification we consider is a failure to account for
zero inflation when the true data were generated according to the Z-NMF model --
though not pursued here, it could be interesting to study robustness of study
conclusions to misspecification in the number of topics or deliberate
contamination.

For the LDA experiment, we vary the number of samples $D \in \{20, 100\}$, the
number of features $V \in \{325, 650\}$, and the total count per sample $N \in
\{1625, 3250, 6500\}$, in order to approximately match dimensions typical in
real microbiome datasets. On the other hand, we fix the number of topics to $K =
2$ and the Dirichlet hyperparameter to $\alpha = \gamma = 1$. For each simulated
data set, we perform estimation using MCMC sampling, VB, and a parametric
bootstrap. In more detail, this bootstrap procedure fits VB to the original
data, simulates $B = 500$ new datasets $X^{\ast}_{b}$ according to the LDA model
using VB-estimated parameters $\{\hat{\theta}^{\ast}_{b},
\hat{\beta}^{\ast}_{b}\}$, and re-estimates parameters
$\{\hat{\theta}^{\ast\ast}_{b}, \hat{\beta}^{\ast\ast}_{b}\}$ on each simulated
data set, again using VB. The motivation for this procedure is the desire to
strike an easily-parallelizable compromise between MCMC sampling, which can be
time consuming but has reliable uncertainty estimates, and VB, which is fast,
but can underestimate uncertainty \citep{wang2005inadequacy}.

\begin{figure}[!p]
  \centering\includegraphics[width=\textwidth]{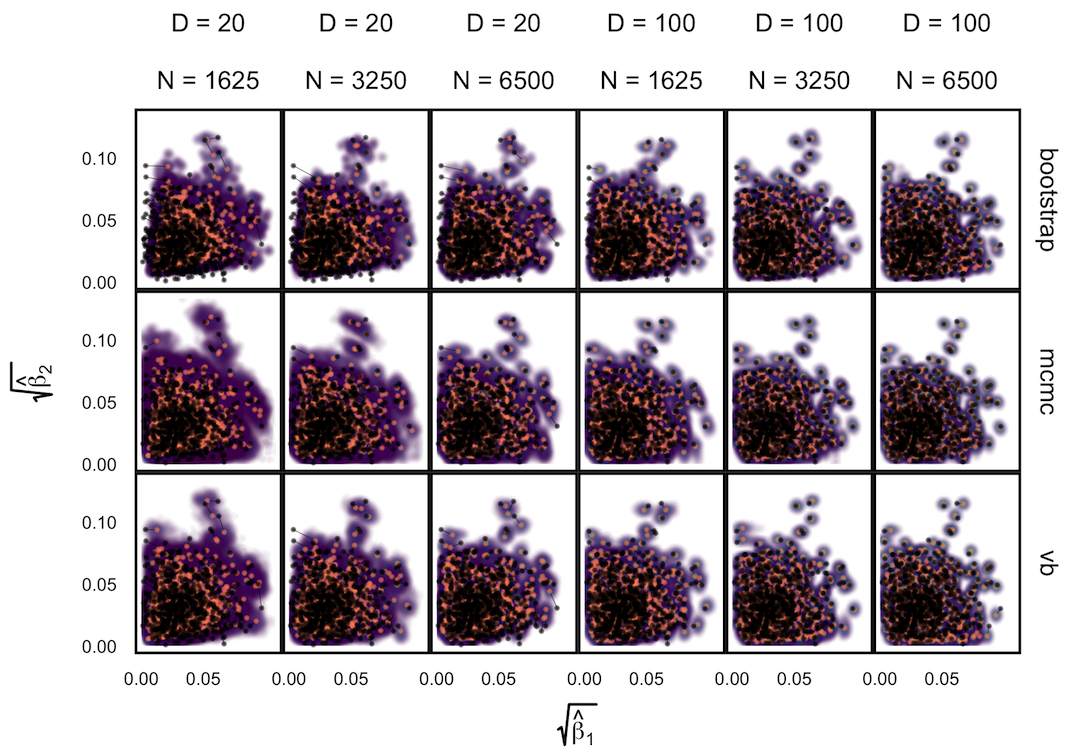}
  \caption{Different inference algorithms for LDA provide comparable posterior
    estimates under the simulation configurations we have considered. Within
    each panel, we display the true pairs $\left(\sqrt{\beta_{v1}},
    \sqrt{\beta_{v2}}\right)$ as black points $v$. The purple clouds are
    posterior samples from the inference procedure, which are given as row
    labels. The posterior medians are orange points that are linked to the true
    $v$. Different columns index different $D$ (top column label) vs. $N$
    (bottom column label) pairs. Here, we have subsetted to $V = 650$ -- the
    corresponding figure for $V = 325$ is given by \ref{fig:beta_contours_v325}.
  }
  \label{fig:beta_contours_v650}
\end{figure}

\begin{figure}[!p]
  \centering
  \includegraphics[width=\textwidth]{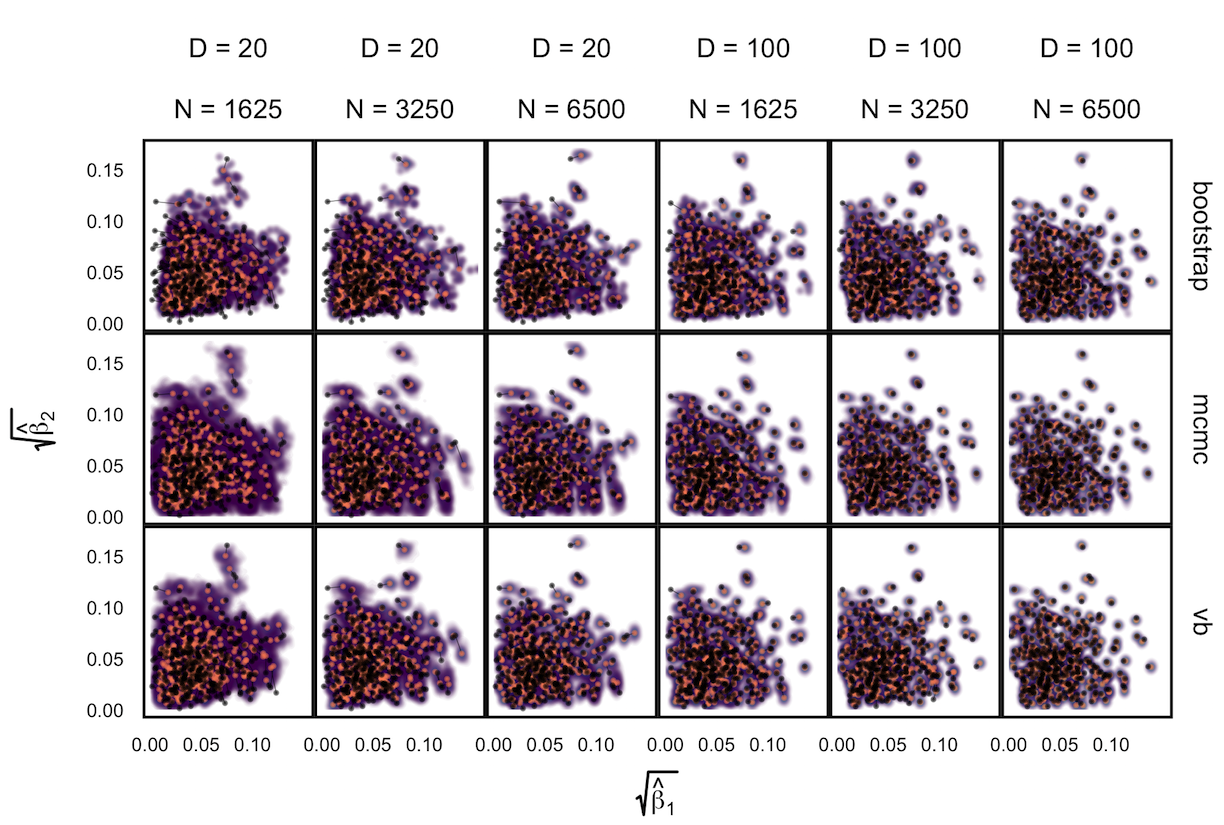}
  \caption{The analog of Figure \ref{fig:beta_contours_v650} in the case that $V
    = 325$. \label{fig:beta_contours_v325}}
\end{figure}

\begin{figure}[!p]
  \centering
  \includegraphics[width=\textwidth]{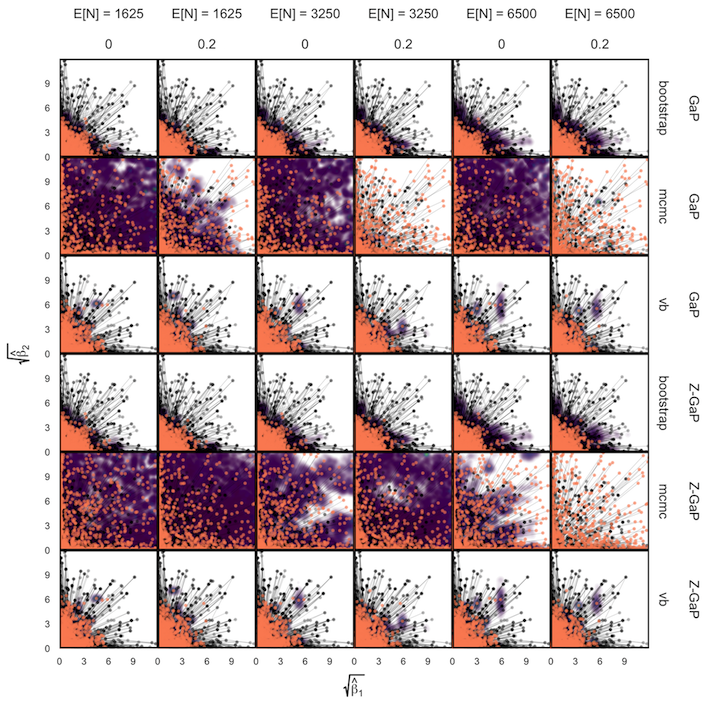}
  \caption{The results of the NMF experiment with $D = 20$ Within each panel, we
    display the true value of $\sqrt{\beta_{v}}$ as black points, while linked
    orange points give associated posterior medians. Note that the axes are
    truncated, and for some panels, the posterior medians all lie outside the
    visible box. Across columns, we vary and $\Earg{N}$. Along rows, we vary the
    assumed model, the inference procedure, and the true $p_{0}$ -- these are
    the three columns of row labels, read from outside
    in.\label{fig:beta_contours_nmf_d20}}
\end{figure}

\begin{figure}[!p]
  \centering
  \includegraphics[width=\textwidth]{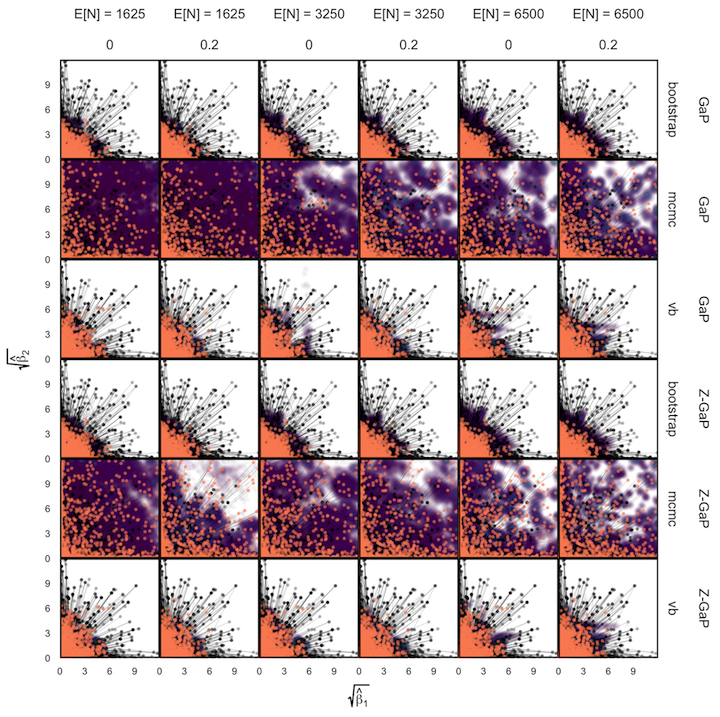}
  \caption{The analog of Figure \ref{fig:beta_contours_nmf_d20} when $D = 100$.
  \label{fig:beta_contours_nmf_d100}}
\end{figure}

Note that, due to the label-switching problem, it is not possible to directly
compare the estimated topics across simulation configurations. To address this
issue, we attempt to identify the permutation that aligns topics across all
experiments. For each experiment, we identify the true-estimated topics pair
with highest correlation, then find the next highest pair among the remaining,
and so forth. Judging from the aligned densities in Figure
\ref{fig:beta_contours_v650} and Supplementary Figures
\ref{fig:beta_contours_v325}, \ref{fig:beta_contours_nmf_d20}, and
\ref{fig:beta_contours_nmf_d100}, this ad-hoc alignment procedure seems
sufficient.

Figure \ref{fig:beta_contours_v650} displays the true and posterior $\beta_{k}$
for the LDA experiment with $V = 650$ features, while varying other simulation
characteristics. Each panel represents a single experimental configuration, with
each axis associated with an underlying topic. Each black point is the
square-root transformed value of feature $v$ across topics:
$\left(\sqrt{\beta_{v1}}, \sqrt{\beta_{v2}}\right)$. The shaded clouds are
sampled posteriors, and the linked orange point give the posterior median for the
associated feature. Across rows, different inferential procedures are compared.
The top row of column labels refers to the total count $N$ within each sample,
while the second refers to the number of samples $D$. The analogous figure when
$V = 325$ is provided in Supplementary Figure \ref{fig:beta_contours_v325}.

As expected, when $D$ increases, the posterior for $\beta$, whose dimension does
not increase with $D$, begins to concentrate around its true value. As the
number of samples or total count within samples increases, the posteriors
further concentrate around the truth. In each of the settings considered, the
three methods seem comparable, suggesting that for microbiome studies of this
approximate scale, VB may be a reasonable choice, considering that it can be run
much more quickly than either MCMC or the bootstrap.

\begin{figure}[!p]
  \centering
  \includegraphics[width=\textwidth]{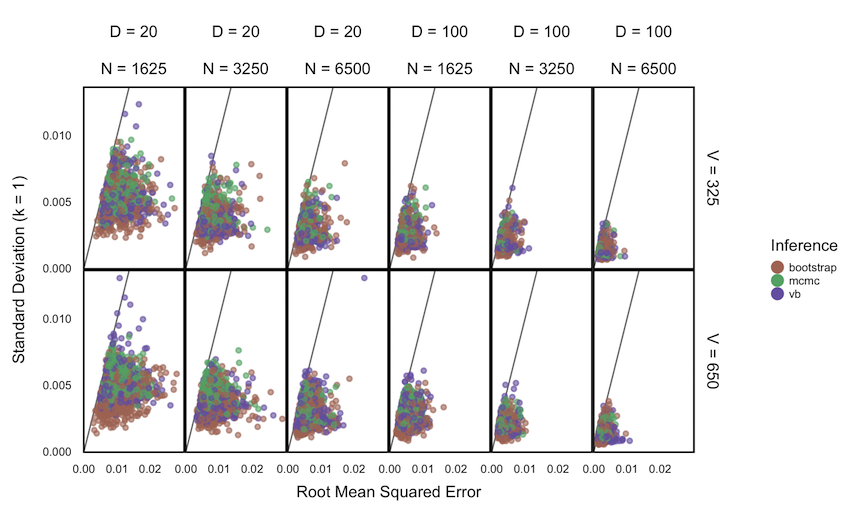}
  \caption{A summary of the errors and uncertainty across models and regimes for
    the $\beta_{vk}$ in Figure \ref{fig:beta_contours_v650} and Supplementary
    Figure \ref{fig:beta_contours_v325}. On the $x$-axis, we plot the difference
    between posterior median and the true value, after having square-root
    transformed. On the $y$-axis, we provide the standard deviation of the
    posterior samples for each $v$, along the first dimension $K = 1$. Columns
    are indexed as in Figure \ref{fig:beta_contours_v650}, but now rows provide
    different values of $V$.
    \label{fig:beta_errors_lda} }
\end{figure}

While the kernel-smoothed posterior densities display the complete results from
this simulation study, a few summary statistics from these densities can
facilitate comparison across models and data generation regimes. The key
features of interest are (1) the distance of the posterior medians for each
parameter from their true values, after alignment, and (2) the concentration of
these posteriors around their medians. Figure \ref{fig:beta_errors_lda}
addresses these questions directly -- the $x$-axis gives the RMSE of the
square-root-transformed posterior medians for the $\beta_{vk}$, for each $v$,
across $k$, and the $y$-axis gives the associated standard deviation, along $K =
1$, for each $v$. As in Figure \ref{fig:beta_contours_v650}, the first row of
column numbers gives the total count $N$ in each sample, and the second gives
$D$, the number of samples. The grey line within each panel is the identity
line, where the error equals one standard deviation.

The drift of points to the bottom left as $N$ and $D$ increase reflects the
improved accuracy and concentration of all inference techniques when the number
of samples and total count within samples increase. Note that few points lie
above the identity standard deviation line -- this suggests that the posteriors
may be misleadingly narrow. Finally, we find that the scale of errors and
concentration in the $V = 325$ and $V = 650$ cases are comparable. Considering
that there is the same number of documents available for estimating each
individual $\beta_{vk}$, this is not surprising.

\begin{figure}[!p]
  \centering
  \includegraphics[width=\textwidth]{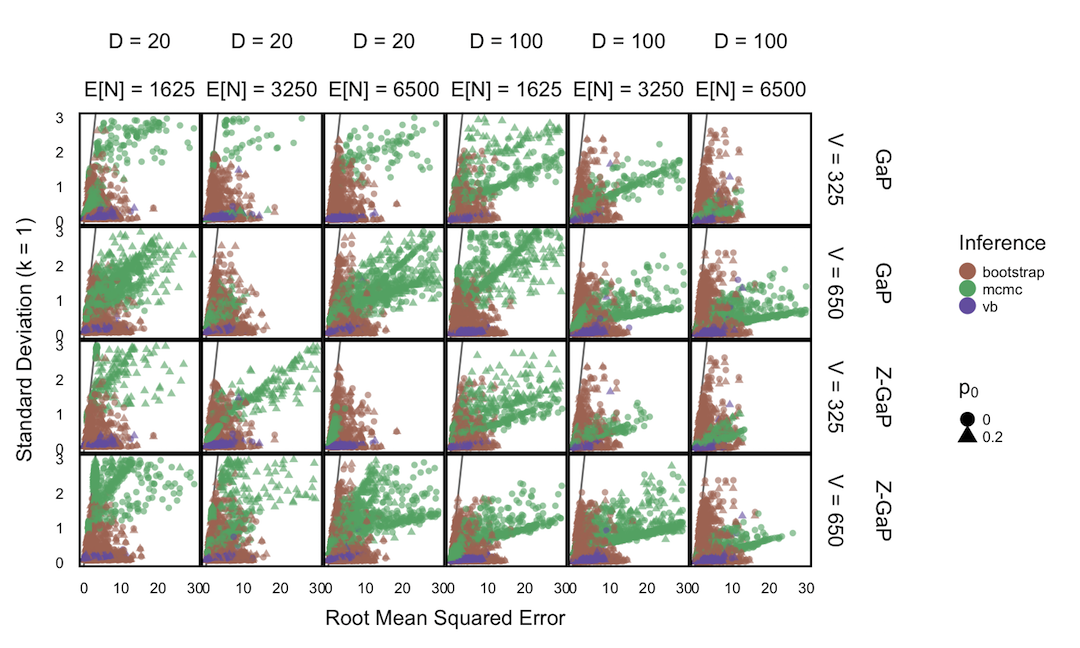}
  \caption{A version of Figure \ref{fig:beta_errors_lda} for the NMF simulation
    experiment. The figures are read similarly, except there are a larger number
    of experimental configurations -- rows now distinguish the assumed model and
    shapes represent the true value of $p_{0}$. Further, while the first row of
    column labels still gives $D$, the second row gives $\Earg{N}$ instead of
    $N$. Note that we also now truncate the $x$ and $y$ axes, and not all points
    are visible. For example, most MCMC samples in the last panel in the second
    row have errors and SDs larger than what is displayed, and so are
    missing. \label{fig:beta_errors_nmf}}
\end{figure}

For the NMF experiment, we use similar parameters, except now we introduce a
zero-inflation probability, $p_{0} \in \{0, 0.2\}$, and have control only over
the expected total count per sample $\Earg{N}$, rather than $N$. When $p_{0} >
0$, we perform inference when the true $p_0$ is provided and also when $p_{0}$
is (incorrectly) assumed to be zero. To modulate $\Earg{N}$, we vary $a_{0}$ and
$b_{0}$ so that $\Earg{N} = \frac{b_{0}}{a_{0}}KV \in \{1625, 3250, 6500\}$,
matching the LDA experiment. As before, we vary the number of samples $D \in
\{20, 100\}$ but fix the number of latent factors $K$ at 2. Figure
\ref{fig:beta_errors_nmf} summarizes the estimation error across regimes, and is
read in the same way as Figure \ref{fig:beta_errors_lda}, with two exceptions,
\begin{enumerate}
\item Instead of providing the fixed values of $N$, we display $\Earg{N}$.
\item We must distinguish between different $p_{0}$ regimes, and whether or not
  the inferential procedure assumed $p_{0}$ was 0 or was given the true $p_{0}$.
  The true value of $p_0$ is marked by the second row of column labels, and
  inference where $p_{0}$ was provided is denoted by the row label Z-GaP, while
  those rows marked GaP assumed $p_{0} = 0$.
\end{enumerate}
The associated posterior density display is given in Supplementary Figures
\ref{fig:beta_contours_nmf_d20} and \ref{fig:beta_contours_nmf_d100}.

The results in Figure \ref{fig:beta_errors_nmf} are more problematic than those
in Figure \ref{fig:beta_errors_lda}. First, the standard errors are no longer
comparable across methods, with those for the MCMC sampling procedure appearing
much larger across many simulation configurations. Second, the error rates are
substantially larger than those in LDA. Examining Supplementary Figures
\ref{fig:beta_contours_nmf_d20} and \ref{fig:beta_contours_nmf_d100}, it appears
that lack of identifiability in the NMF model may be affecting our ability to
evaluate the resulting fits. The only way to distinguish between models
$\left(c\Theta\right)\left(\frac{1}{c} B\right)^{T}$ and $\Theta B^{T}$ is
through the prior, and the prior may not be strong enough to guide inference to
the right scaling.

Further, examining Supplementary Figure \ref{fig:beta_contours_nmf_d20}, we note
many cases where estimation fails to converge or concentrates around the
incorrect value, though this may be related to rescaling. Generally, we do not
find a direct probabilistic-programming implementation of the GaP or Z-GaP
models reliable in the regimes under consideration, even when $p_{0}$ is known.

Next we consider a simulation experiment involving the unigram model. We use the
same configurations of $D$, $V$, and $N$ as in the LDA and NMF experiments
above. The parameter controlling the rate of evolution on the simplex,
$\sigma_{0}^2$, is set to 1 throughout. During inference, $\sigma_0^2$ is not
assumed known, and is instead modeled as an $\text{InvGamma}\left(1, 1\right)$
across all simulation settings.

\begin{figure}[ht]
  \centering
  \includegraphics[width=\textwidth]{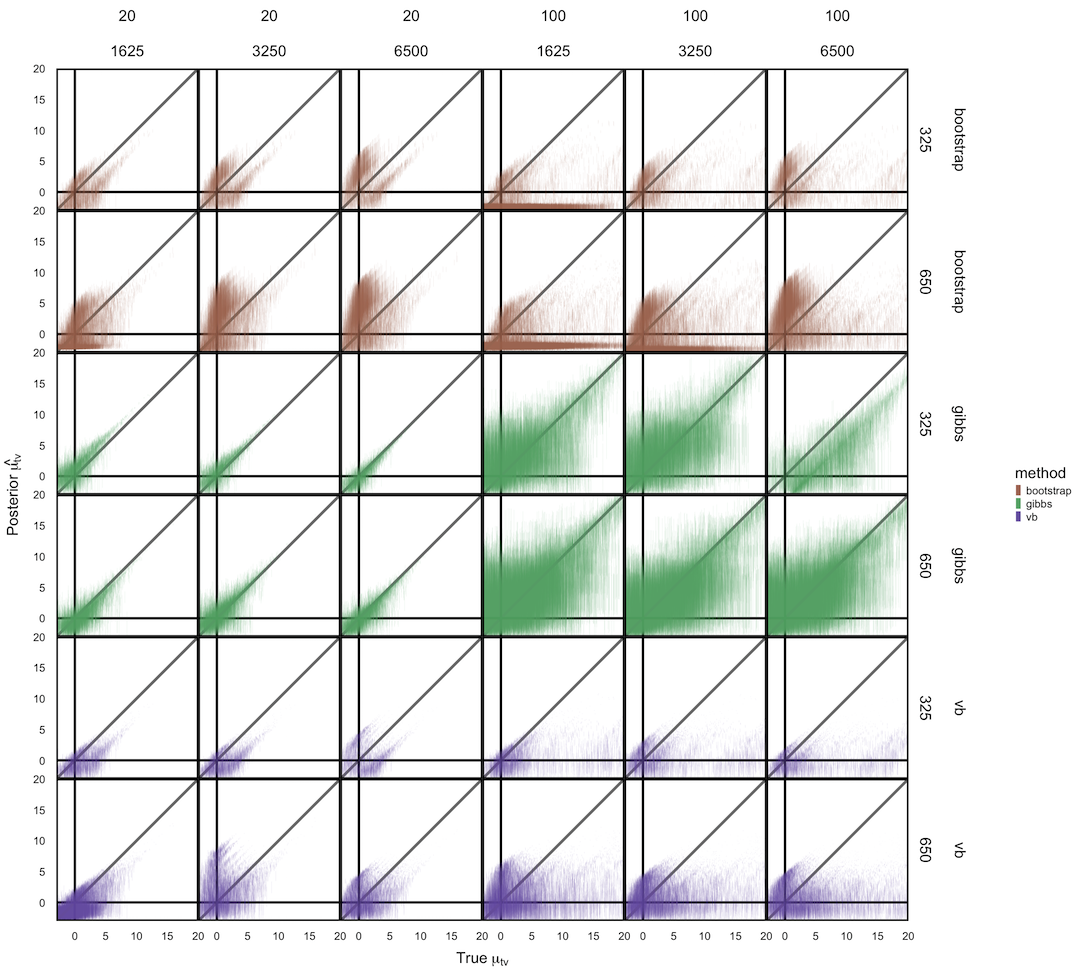}
  \caption{
    A comparison of the posterior $p\left(\mu_{tv} \vert x\right)$ to the known
    underlying $\mu_{tv}$, in the unigram simulation experiment. The $x$-axis
    for each interval corresponds to the true $\mu_{tv}$ for one species at one
    timepoint, while the vertical intervals cover the 25\% to 75\% quantiles of
    samples from the posterior $p\left(\mu_{tv} \vert x\right)$. Different
    panels distinguish between configurations of $D$, $N$, $V$, and posterior
    sampling schemes. Posteriors from MCMC sampling seem to correctly recover
    the true underlying $\mu_{tv}$, while discrepancies arise for both VB and
    the bootstrap.
    \label{fig:mu_intervals} }
\end{figure}

In the NMF and LDA experiments, we could visualize fitted posteriors over
$\theta_i$ and $\beta_j$ as two-dimensional smoothed scatterplots, since we set
$K = 2$. In contrast, this unigram experiment involves a $V$-dimensional
parameter $\mu_t$ evolving over time -- this precludes any direct analog to
Figure \ref{fig:beta_contours_v650} or Supplemental Figure
\ref{fig:beta_contours_nmf_d20}. Instead, in Figure \ref{fig:mu_intervals}, we
visualize the true $\mu_{tv}$ against posterior intervals for the
one-dimensional $p\left(\mu_{tv} \vert x\right)$, across all configurations of
simulation parameters.

Evidently, MCMC provides accurate posterior estimates, while VB and the
bootstrap estimates deviate from the true $\mu_{tv}$. First, when the true
$\mu_{tv}$ is small small, the posterior estimates from VB and the bootstrap are
too large. This is not as much a reason for concern as it might appear at first,
however, as the parameters $\mu_t$ are passed through a softmax before being
mapped to probabilities. After this transformation, the difference between, say,
$\mu_{tv} = 10^{-2}$ and $\mu_{tv} = 10^{-5}$ is relatively unimportant. Second,
posterior estimates for $\mu_{tv}$ for large positive values of the parameter
tend to be biased downards, to a degree that can't be simply explained as an
effect of the prior, as such a strong bias is not present among the MCMC-sampled
posteriors.

To simplify the comparison between methods, we can also display the unigram
analog of Figure \ref{fig:beta_errors_lda} -- this is given in Supplementary
Figure \ref{fig:mu_errors_unigram}. This display confirms our earlier
observation, that posterior MCMC samples seem more reliable than either those
from VB or the bootstrap, when using generic probabilistic programming for the
Dynamic Unigram model, at least in problems with the dimensions we have
considered.

\section{Data Analysis}
\label{sec:data_analysis}

In applying probabilistic methods to microbiome data analysis, we concentrate on
two questions,
\begin{enumerate}
\item Do these models fit the data well, and what techniques are available for
  performing this evaluation?
\item Supposing these models fit well, do they lend themselves to informative
  summaries of the original data?
\end{enumerate}

To develop answers to these questions, we reanalyze the data of
\cite{dethlefsen2011incomplete}, a study of bacterial dynamics in response to
antibiotic treatment. This study monitored the microbiomes of three patients
over ten months, with two antibiotics time courses introduced in between, in
order to study the effect of antibiotic perturbations within the context of
natural long-term dynamics. By applying Principal Components Analysis, the study
concluded that antibiotics cause substantial changes in short-term community
composition, with certain species being substantially more resilient than
others, and also detected long-term effects in one patient. The purpose of our
case study is to compare these conclusions with those obtained through
probabilistic latent variable models.

Variation in bacterial signatures tends to be dominated by strong inter-subject
effects \citep{eckburg2005diversity}, and with only three subjects, there is
little reason for a model which clusters across subjects. Hence, we choose to
study one individual at a time. In this section, we focus on Subject F, who had
been reported to exhibit incomplete recovery of the pre-antibiotic treatment
bacterial community. However, analogous figures for the other two subjects are
available as Supplementary Figures \ref{fig:antibiotics_lda_theta_D} through
\ref{fig:antibiotics_lda_beta_E}. There are 2582 distinct species across these
samples, and we perform no pre-filtering, though most species are present in
very low numbers.

It is possible that future studies may consider a similar design, but involve
many more subjects. In this case, it would be possible to do better than model
each subject separately. The simplest way to share information across subjects
would be to still have separate topics $\beta_{k}^{s}$ for each individual $s$,
but to place a common prior across all $\left(\beta_{k}^{s}\right)_{k, s}$. This
approach has the disadvantage that topics are not directly comparable across
subjects. An alternative that mitigates this issue would be to share topics
$\left(\beta_{k}\right)_{k}$ across all individuals, choosing $K$ to be larger
than would be appropriate for any single subject. In this case, it would be
expected that some $\theta_{ik}$ would be near zero for some topics for all
samples within a subject -- that subject may never have samples drawn from a
particular topic. It is possible to directly incorporate this behavior, in which
some but not all topics are shared across subjects, into the model form by
considering a hierarchical Dirichlet prior, rather than a standard one
\citep{wallach2006topic, teh2005sharing}.

In this case study, we focus on LDA and the Dynamic Unigram model. A similar
study using GaP and Z-GaP is omitted, in light of the difficulties in estimation
during the simulation studies. Throughout, we apply VB, though considering the
results of the simulation study, we exercise caution when interpreting estimated
uncertainties. We set $K = 4$, based on the heuristic that a larger $K$ would be
less meaningful, since there are only 56 timepoints. In cases where it is of
interest to choose $K$ automatically, it would be possible to evaluate the
likelihood of fitted models on a hold-out set, for a range of values of $K$, by
fixing $\left(\hat{\beta}_{k}\right)_{k = 1}^{K}$ and estimating new
$\hat{\theta}_{i}$ for the hold-out samples \citep{blei2003latent}. Indeed, this
process reflects an advantage of probabilistic modeling: the Bayesian Occam's
razor ensures that the posterior probability of a model won't always go up as
$K$ increases -- contrast this with ordinary $K$-means, where tailored methods
like the Gap or Silhouette statistics must be studied
\citep{rasmussen2001occam, tibshirani2001estimating, kaufman2009finding}.
Alternatively, the proposals of \citep{wallach2009evaluation} could be carried
out. However, in this study, we guide the choice of $K$ based on what we find
most useful for scientific interpretation, rather than that which necessarily
gives the best fit over the entire population.

Note that we plot the fitted probabilities on a logit scale -- for a raw vector
of probabilities $\mathbf{p} = \left(p_{1}, \dots, p_{D}\right) \in
\simplex^{D - 1}$, we plot $g\left(\mathbf{p}\right) := \left(\log p_{1} -
\overline{\log \mathbf{p}}, \dots, \log p_{K} - \overline{\log
  \mathbf{p}}\right)$, which are similar to log-odds, but centered according to
the average log probability, rather than any reference class.

Both LDA and the Dynamic Unigram model are more computationally intensive than
more common approaches, like principal components analysis (PCA) or
multidimensional scaling (MDS), but are still amenable to routine application.
For example, in the case studies below, LDA and the Dynamic Unigram model can
both be compiled and run on a standard laptop\footnote{1.4GHz Intel Core i5
  processor and 4GB RAM} in 5.62 and 30.0 minutes, respectively -- the longer
runtime of the Dynamic Unigram model is due to its larger number of parameters.
Further, both approaches can be scaled to much larger datasets without requiring
substantially more memory (but requiring somewhat longer training time) by
applying stochastic variational inference \citep{hoffman2013stochastic,
  kucukelbir2015automatic}.

\subsection{Latent Dirichlet Allocation}
\label{sec:antibiotics_lda}

\begin{figure}[!p]
  \centering\includegraphics[width=\textwidth]{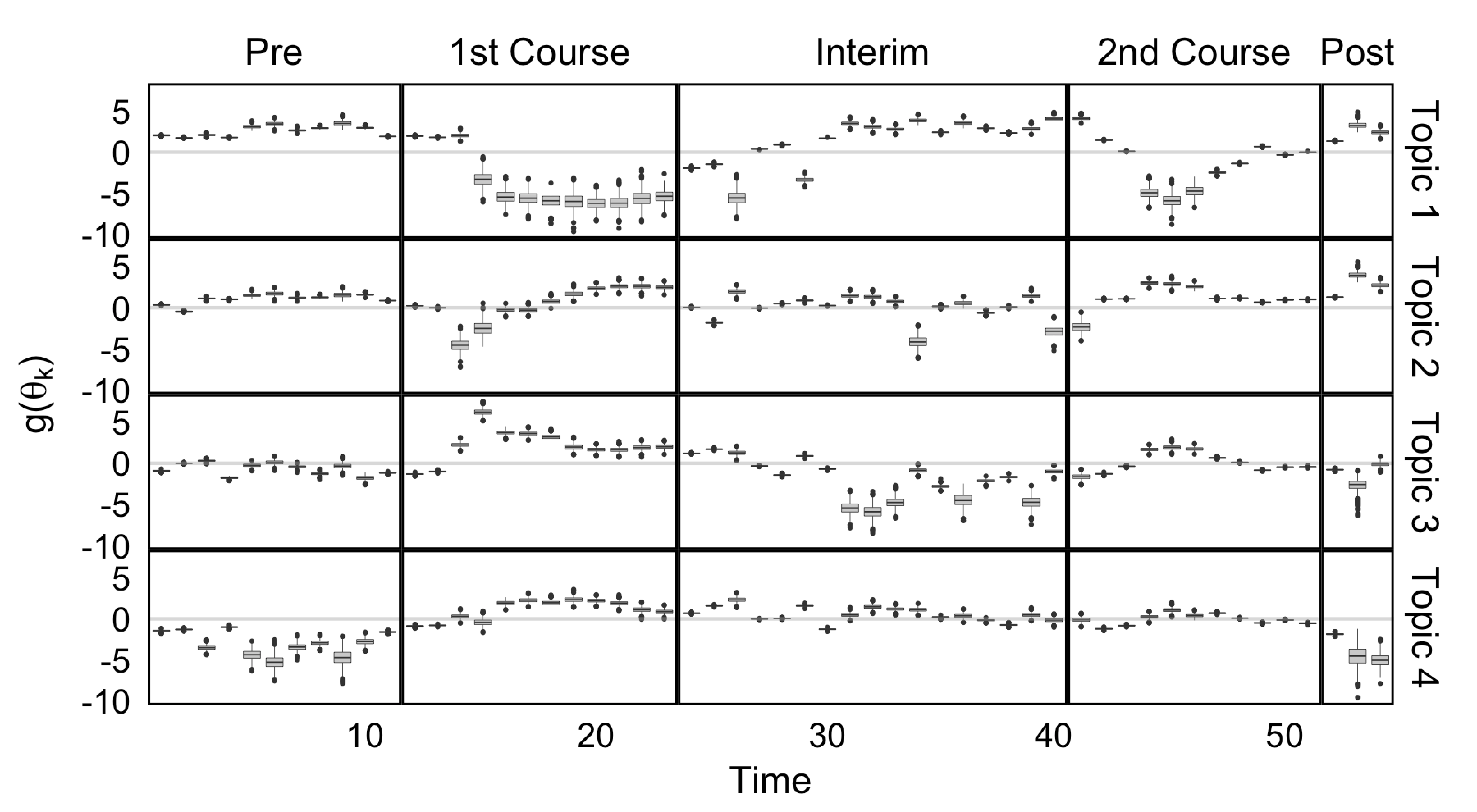}
  \caption{Boxplots represent approximate posteriors for estimated mixture
    memberships $\theta_{d}$, and their evolution over time. That is, each row
    of panels provides a different sequence of $\theta_{dk}$ for a single $k$,
    and different columns distinguish different phases of sampling. Note that
    the $y$-axis is on the $g$-scale, which is defined as a translated logit,
    $g\left(\mathbf{p}\right) := \left(\log p_{1} - \overline{\log \mathbf{p}},
    \dots,\log p_{K} - \overline{\log \mathbf{p}}\right)$. Note that the first
    and second antibiotic time courses result in meaningful shifts in these
    sequences, and that there appear to be long-term effects of treatment among
    bacteria in Topic 3. \label{fig:antibiotics_lda_theta}}
\end{figure}

\begin{figure}[!p]
  \centering\includegraphics[width=\textwidth]{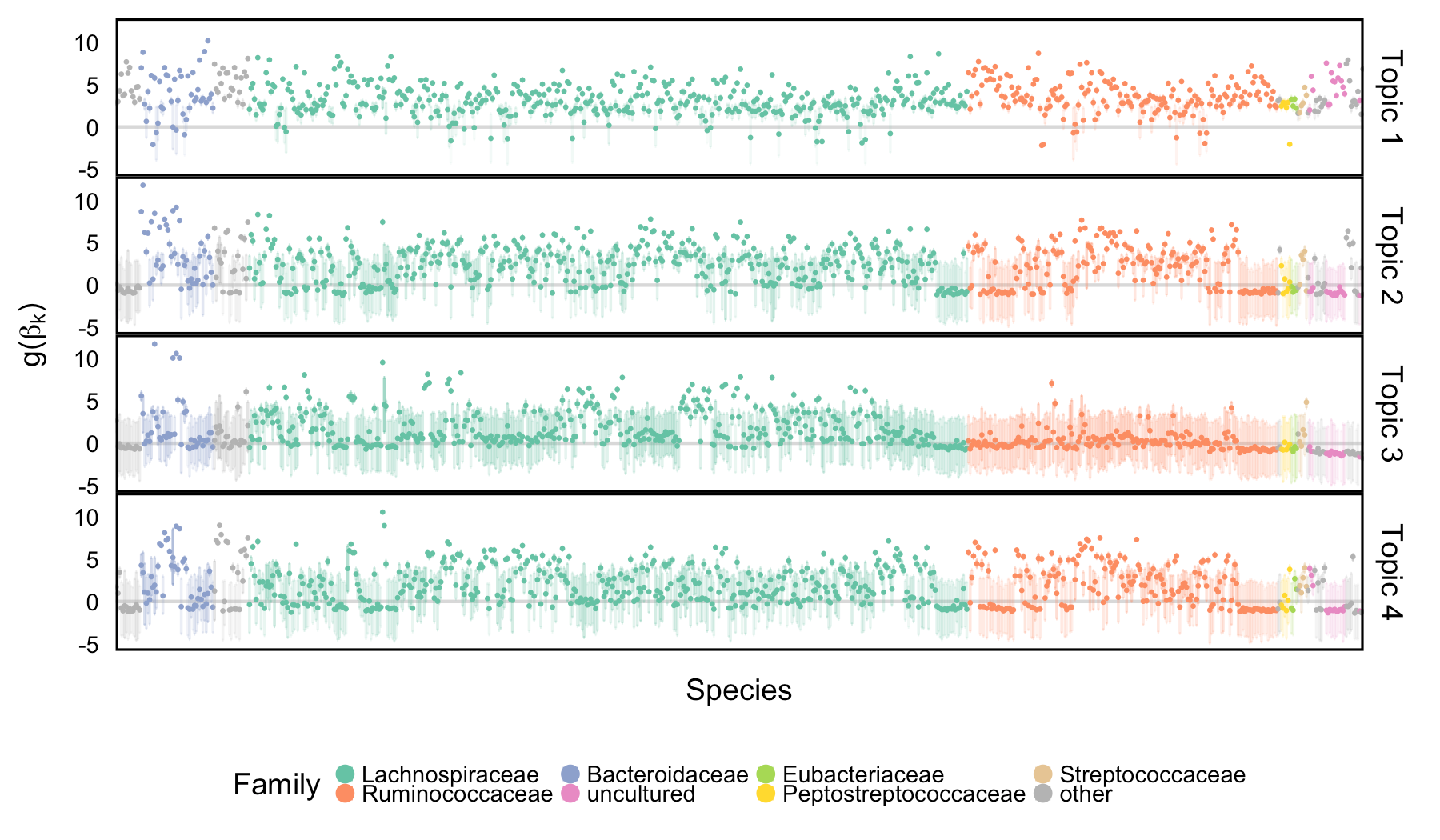}
  \caption{Each credible interval describes an approximate posterior for one
    $\beta_{vk}$. Coupled with Figure \ref{fig:antibiotics_lda_theta}, this
    guides the interpretation of which bacterial taxa are more or less prevalent
    during antibiotic treatments. Each row of panels corresponds to one of the
    four topics, the $x$-axis indexes species, sorted according to phylogenetic
    relatedness, and the $y$-axis give transformed values of the species probability
    under that topic. Only the 750 most abundant species are shown. Note the
    disappearance of otherwise abundant species within Topics 2, 4, and to some
    extent, 1.} \label{fig:antibiotics_lda_beta}
\end{figure}

\begin{figure}[!p]
  \centering\includegraphics[width=\textwidth]{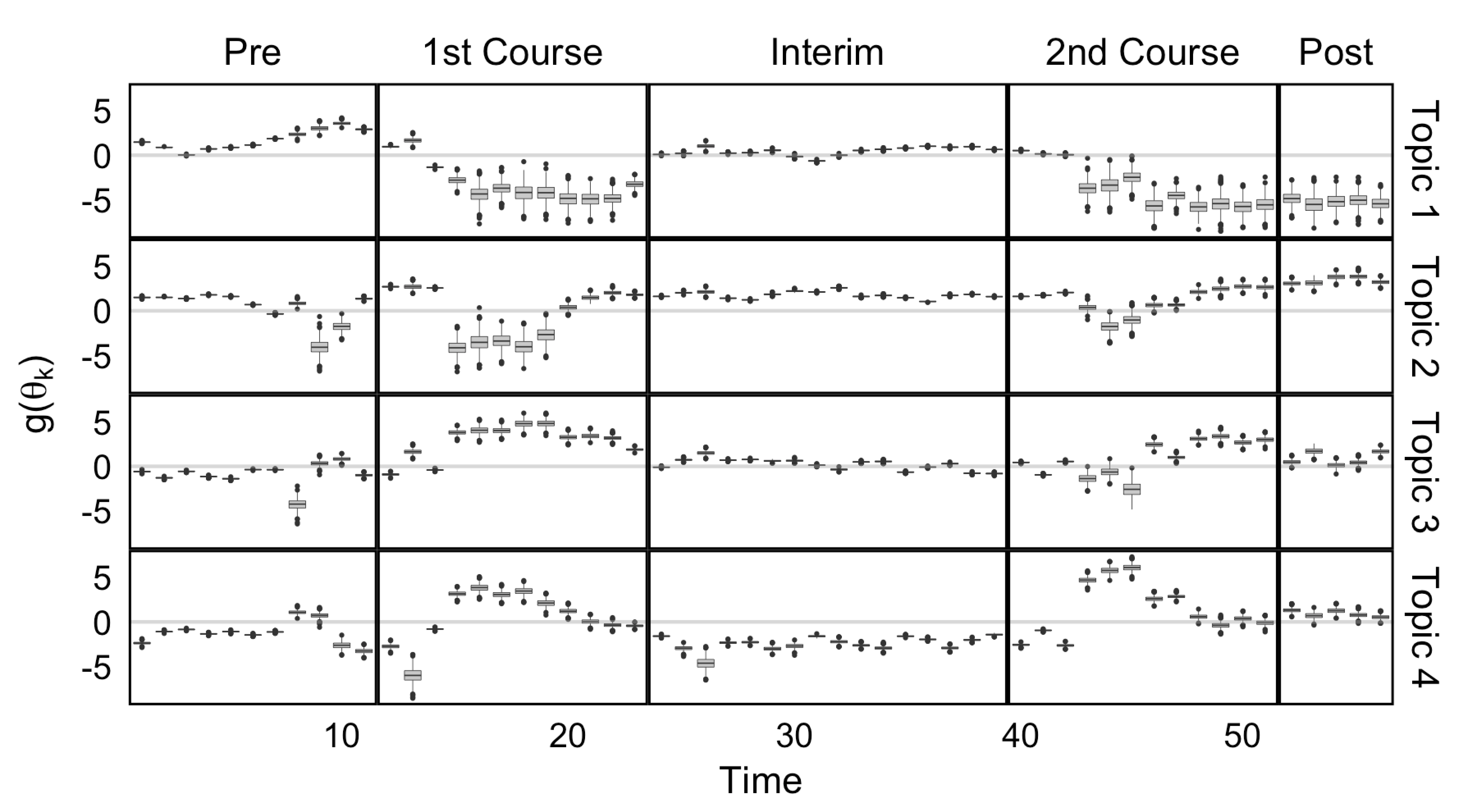}
  \caption{The analog of Figure \ref{fig:antibiotics_lda_theta} for Subject
    D. \label{fig:antibiotics_lda_theta_D}}
\end{figure}

\begin{figure}[!p]
  \centering\includegraphics[width=\textwidth]{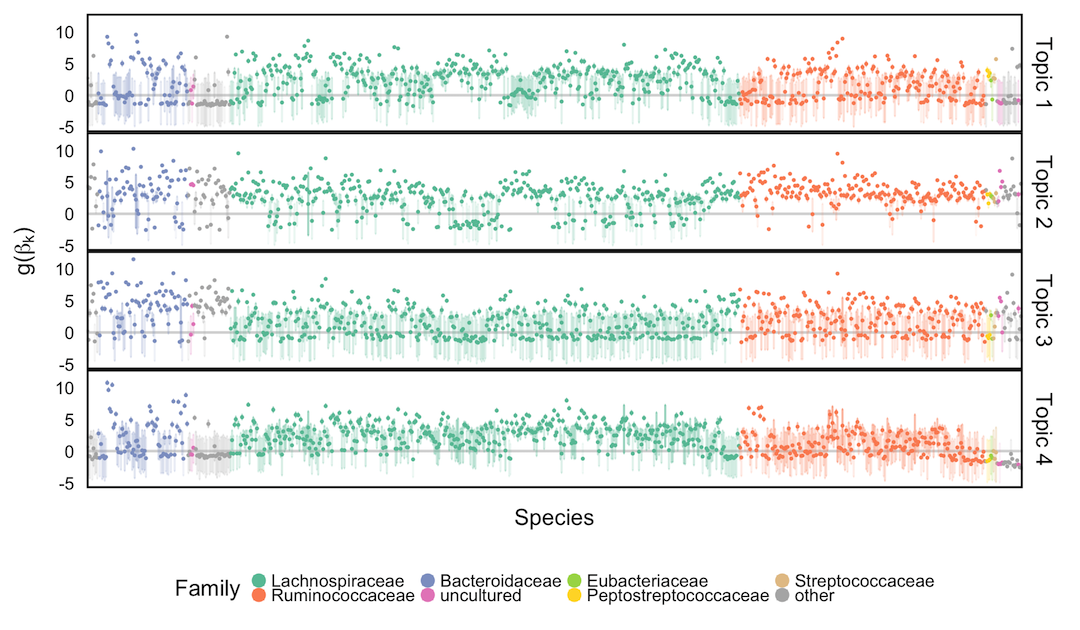}
  \caption{The analog of Figure \ref{fig:antibiotics_lda_beta} for Subject E. \label{fig:antibiotics_lda_beta_E}}
\end{figure}

The fitted parameter values are summarized in Figures
\ref{fig:antibiotics_lda_theta} and \ref{fig:antibiotics_lda_beta}.
In Figure \ref{fig:antibiotics_lda_theta}, rows represent topics, the $x$-axis
represents time, and the $y$-axis gives the boxplots of posterior quantiles for
each $\theta_{dk}$.

This figure draws attention to the two antibiotic time courses, which took place
between days 12-23 and 41-51. Topic 1 seems to become prominent during the
interim period between the two time courses, suggesting that this is the
bacterial community that fills in the niches left empty during the first time
course. Conversely, Topic 3 seems to represent those bacteria that were present
initially but are eliminated during the first time course, though there is a
hint of a recovery at the end of sampling. This topic seems most closely related
to the finding reported in \cite{dethlefsen2011incomplete} that Subject F
experienced long-term antibiotic effects on bacterial community composition.
Topics 2 and 4 seem to be overrepresented during the antibiotic treatments.
Topic 2 is elevated immediately after the antibiotic treatment. Topic 4 seems to
summarize those bacteria that are initially negatively impacted by the
antibiotic treatments, but which recover relatively quickly and have higher
abundance at the end of the trial than at the beginning. The fact that Topics 2
and 4 are learned from the data without enforcing any temporal evolution
structure on the samples suggests that under an antibiotic perturbation, the
system exhibits differential recovery.

To interpret these topics in terms of their bacterial community fingerprint, we
study the estimated topic distributions $\beta_{k}$. This is displayed in Figure
\ref{fig:antibiotics_lda_beta}. The four rows correspond to the $K = 4$
estimated topics. Within a row, we show 95\% credible intervals associated with
the posterior samples for each bacterium. Different colors identify different
taxonomic families, and bacteria are sorted according to evolutionary
relatedness. For clarity, we have displayed only the 750 most abundant bacteria.

Considering the mixture probabilities in Figure \ref{fig:antibiotics_lda_theta},
those bacteria with large probabilities in the second and fourth rows of Figure
\ref{fig:antibiotics_lda_beta} constitute a large fraction of the samples taken
during antibiotics time courses, reflecting those whose abundances increase
rapidly (Topic 2) or are initially drop but then increase gradually (Topic 4)
during the antibiotics time courses. These distributions are relatively more
concentrated on a small subset of bacteria with high probabilities, reflected by
the drop in logitted probabilities far below zero. This corresponds to a
decrease in community diversity during antibiotic time courses. We also note
groups of neighboring bacteria with similarly elevated topic probabilities. It
is encouraging that, even without specifying smoothness along the phylogenetic
tree in the prior for the $\beta_{k}$s, such smoothness emerges in the fitted
model.

\begin{figure}[!p]
  \centering\includegraphics[width=\textwidth]{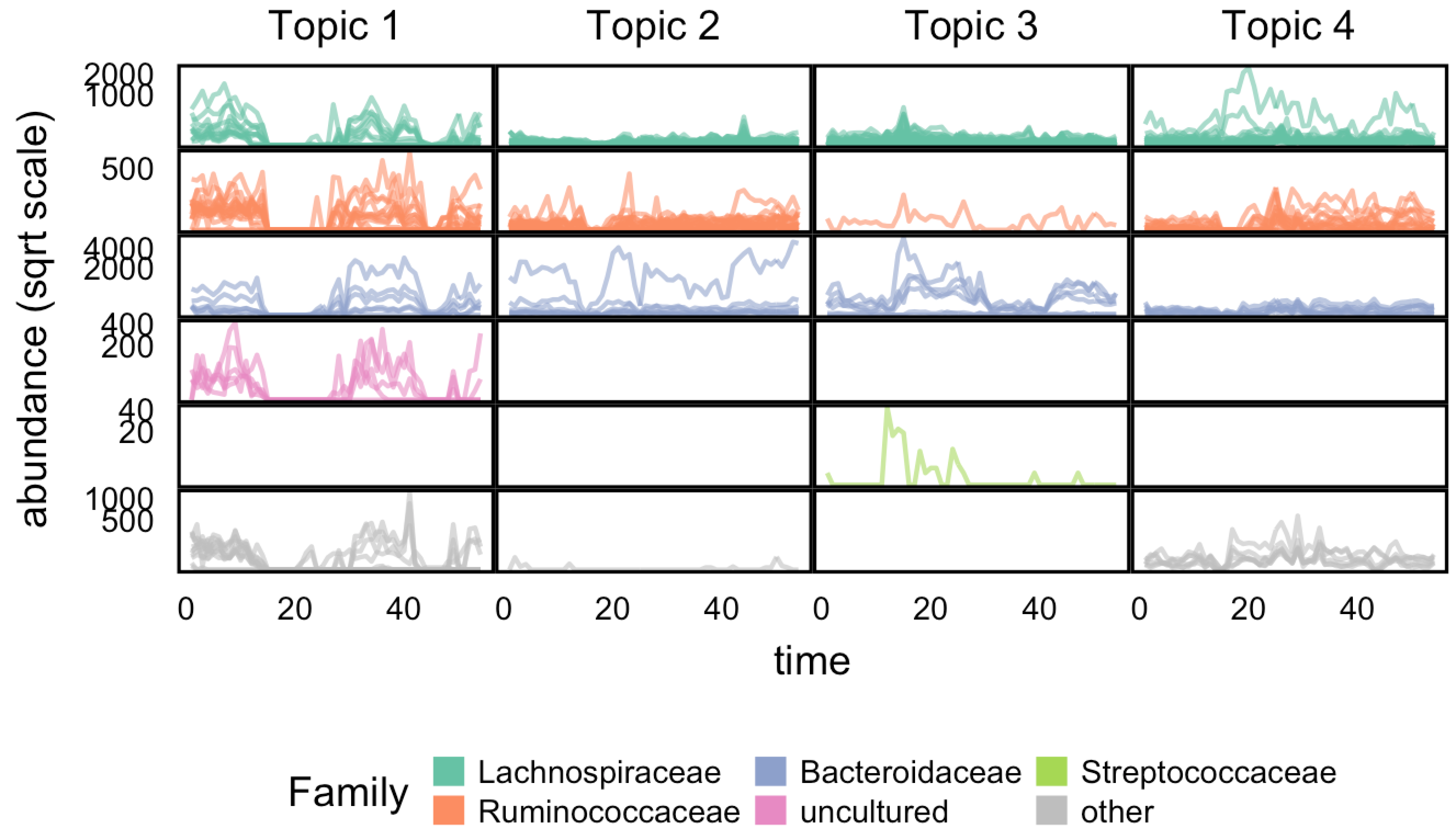}
  \caption{Individual species that have been identified as prototypical of
    single topics are highlighted here, to provide a characterization of topics
    linked to the original bacterial abundance data. Rows and columns of panels
    correspond to taxonomic families and estimated topics,
    respectively, and each trajectory gives the abundance of a single species
    over time. Note that the $y$-axis is on a square root scale. This view
    provides more evidence for the interpretation of the four topics as (1)
    decreased abundance during antibiotic time courses (2) increased abundance during
    antibiotic time courses (3) disappearance after first time course, and (4)
    delayed but sustained increase after first time
    course. \label{fig:topic_prototypes}}
\end{figure}

One disadvantage of plotting the posteriors for the $\beta_{k}$ in this way is
that it is difficult to determine the identify of any particular bacterial
species associated with a specific credible interval. It also hides any
interesting variation that may be occurring among species not among the 750 most
abundant. To address these issues, we have designed an interactive version,
available online at
\url{https://statweb.stanford.edu/~kriss1/microbiome_plvm/vis.html}.
Alternatively,
individual species that are primarily assigned to individual topics can be used
to characterize the topics. For example, in Figure \ref{fig:topic_prototypes} we
have screened all species for those that seem primarily associated with a single
topic, and plotted them along with the topic for they are representative.
Specifically, representatives of the $k^{th}$ topic were found by choosing the
50 species $v$ with the largest values of $\beta_{kv} - \sum_{k^{\prime} \neq k}
\beta_{k^{\prime} v}$.

\begin{figure}[!p]
  \centering
  \includegraphics[width=\textwidth]{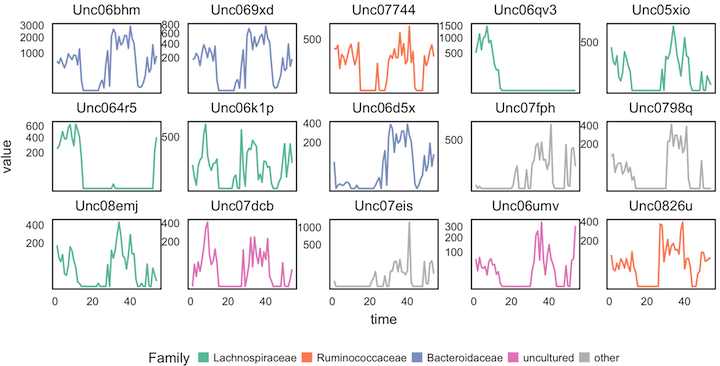}
  \caption{Rather than displaying all representative species together, as in
    Figure \ref{fig:topic_prototypes}, we can sort species according to how
    representative they are of an individual topic. Here, the 15 species most
    strongly associated with Topic 1 are given. The panels are to be read from
    left to right and from top to bottom, to go in decreasing value of
    association. Note that the $y$-axis is on a square root
    scale. \label{fig:species_prototypes_1} }
\end{figure}

\begin{figure}[!p]
  \centering
  \includegraphics[width=\textwidth]{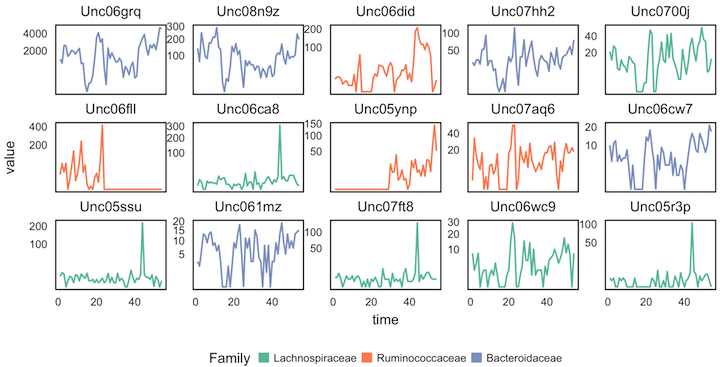}
  \caption{The analog of Figure \ref{fig:species_prototypes_1} for Topic
    2.\label{fig:species_prototypes_2}. }
\end{figure}

\begin{figure}[!p]
  \centering
  \includegraphics[width=\textwidth]{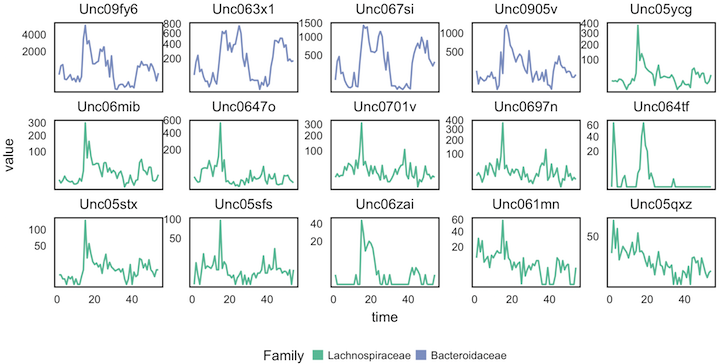}
  \caption{The analog of Figure \ref{fig:species_prototypes_1} for Topic
    3. \label{fig:species_prototypes_3} }
\end{figure}

\begin{figure}[!p]
  \centering
  \includegraphics[width=\textwidth]{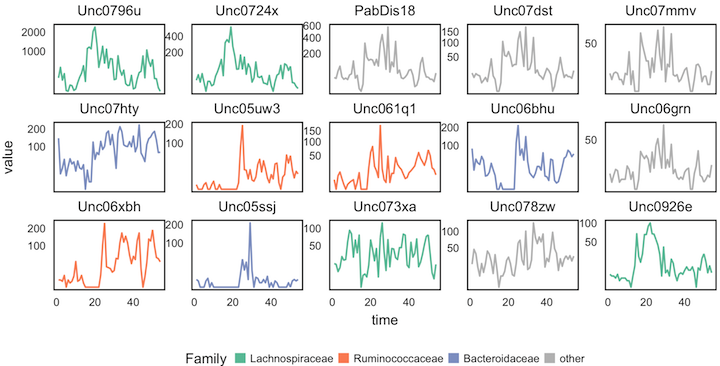}
  \caption{The analog of Figure \ref{fig:species_prototypes_1} for Topic
    4. \label{fig:species_prototypes_4} }
\end{figure}

This allows an interpretation of the topics in terms of the raw data, and the
display reinforces the findings of Figures \ref{fig:antibiotics_lda_theta} and
\ref{fig:antibiotics_lda_beta}. In particular, this view makes clear how much
variation there is in species abundances within taxonomic families -- compare
Topics 1 and 2 among representative Lachnospiraceae and Bacteroides, for
example. Instead of displaying all species together, Supplementary Figures
\ref{fig:species_prototypes_1} through \ref{fig:species_prototypes_4} sort
individual species by the topic representativeness measure.

Similar approaches can be employed to identify taxa that disproportionately
contain species with high membership in particular topics. For example, we can
compute the average topic representativeness statistic defined above across all
species within a taxonomic family in order to characterize that family. Families
that are highly associated with individual topics are displayed in Supplementary
Figures \ref{fig:uneven_taxa_ordered} and \ref{fig:uneven_taxa_facet}.

\subsection{Dynamic Unigram model}
\label{sec:antibiotics_unigram}

\begin{figure}[!p]
  \centering
  \includegraphics[width=\textwidth]{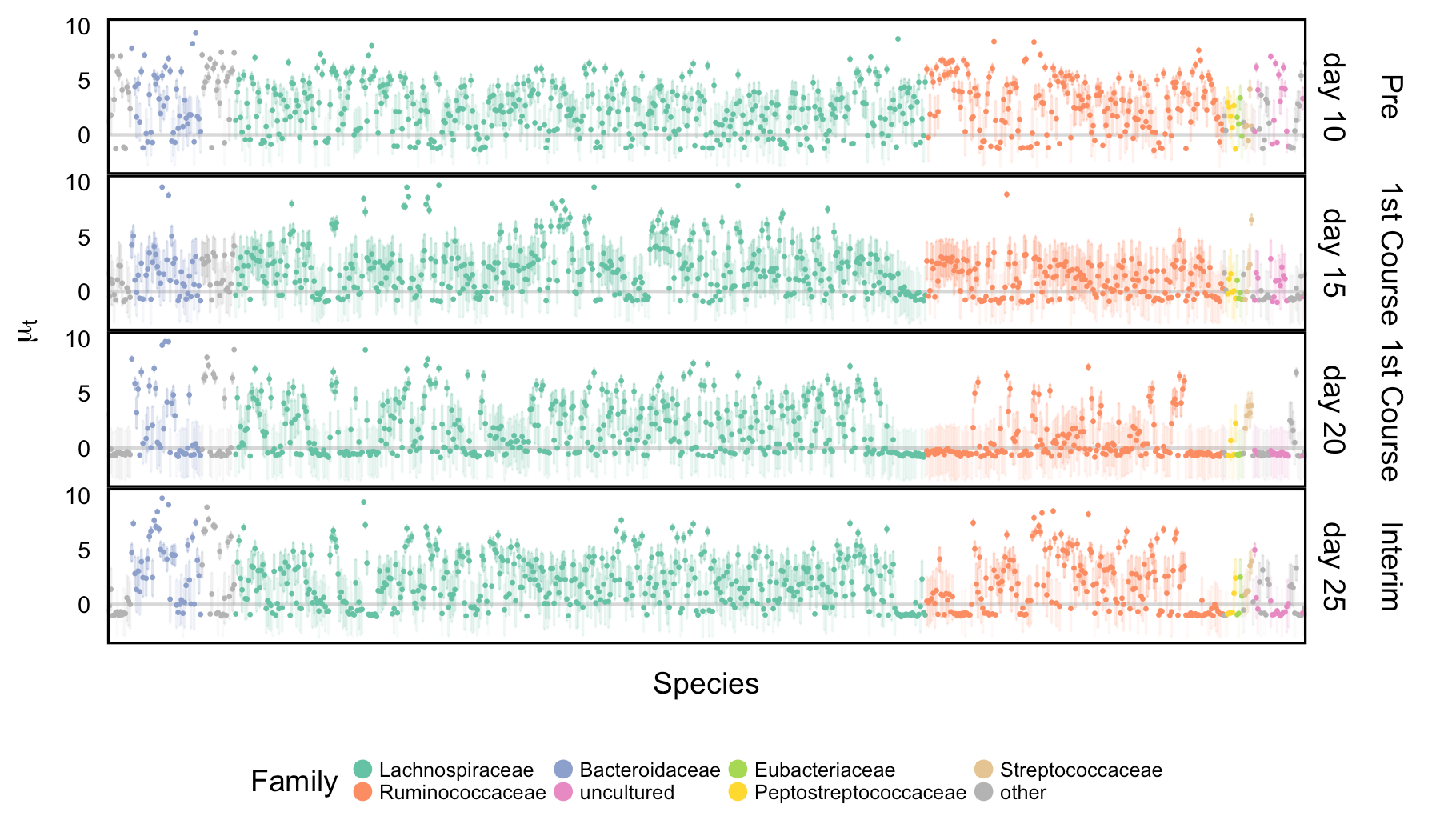}
  \caption{Each posterior credible interval refers to one $\mu_{vt}$. The rows
    are a subset of times $t$ around the first antibiotic time course. The first
    row corresponds to a timepoint from before the treatment, the middle two
    from during the antibiotics time course, and the bottom from after the time
    course was stopped. Otherwise, this display is read in the same way as
    Figure \ref{fig:antibiotics_lda_beta}. This view provides one way of
    smoothing abundance time series, to see how different species respond to
    antibiotic treatment. \label{fig:antibiotics_unigram_theta} }
\end{figure}

While we can interpret the LDA-estimated $\hat{\theta}_{d}$ according to their
temporal context, this information was never directly provided to the algorithm.
In contrast, we can apply the Dynamic Unigram model to the same data, which
explicitly models temporal evolution. Unlike LDA, however, this model does not
seek latent mixture structure. Our primary results are displayed in Figure
\ref{fig:antibiotics_unigram_theta}.

Each row in this figure is interpreted similarly to a row in Figure
\ref{fig:antibiotics_lda_beta}, except now they correspond to estimated
proportions over time $g\left(\mu_{t}\right)$ rather than transformed topics
$g\left(\beta_{k}\right)$. Only four of the 54 total timepoints
is displayed, highlighting a time window around the first antibiotic time
course. Such a display implicitly assumes a relatively smooth interpolation
between timepoints, which is enforced by the form of the Dynamic Unigram model.

This analysis yields conclusions similar to those obtained through LDA, though
reaching them requires somewhat more effort. For example, on day 10, most
$g\left(\mu_{tv}\right)$'s have most of their mass concentrated above zero, and
not many are positioned exceptionally far form the bulk. This is consistent with
higher community diversity before the antibiotic time course. On the other hand,
at time 15, one day after the time course began, most species have quantiles
lower than zero, while a few are positioned much higher than the rest. This
corresponds to a less diverse community, whose membership is concentrated on
those species with outlying intervals. This decrease in diversity seems most
profound at time 20; by time 25, during the interim, much of the community seems
to have recovered.

Further, while we continue to see differential recovery across bacteria, the
effect is not as obvious as in LDA, where this effect was decomposed across
topics. For example, many subintervals of Lachnospiraceae seem to return to
their pre-antibiotics levels by time 25, while the Ruminococceae continue to
have low values of $\mu_{tv}$.

\subsection{Posterior Predictive Checks}
\label{sec:antibiotics_ppc}

While we have found both LDA and the Dynamic Unigram model qualitatively useful,
it is still important to seek more formal diagnostics of model fit. Here, we
consider several posterior predictive checks, as explained in Section
\ref{sec:ppc_overview}.

\begin{figure}[!p]
  \centering
  \includegraphics[width=\textwidth]{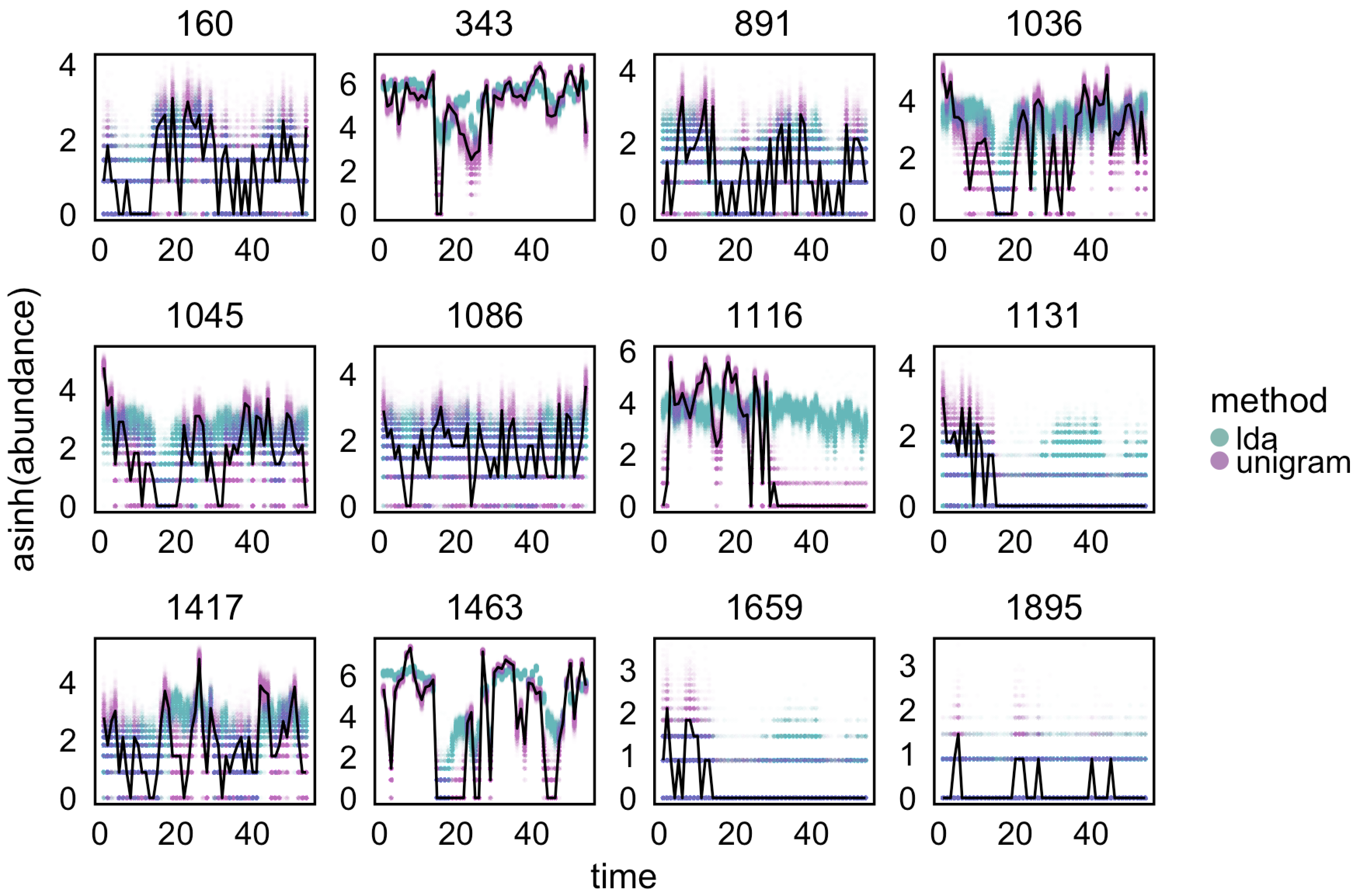}
  \caption{We can visualize the simulated time series for a subset of species
    and compare them with the observed ones, as a posterior check. Each panel
    represents one species. The black lines represent the observed
    $\asinh$-transformed abundances for subject F over time. The blue and purple
    dots give the posterior predictive realizations for these species over time,
    according to LDA and the Dynamic Unigram model, respectively.
    \label{fig:antibiotics_posterior_ts}}
\end{figure}

To this end, in Figure \ref{fig:antibiotics_posterior_ts}, we plot observed time
series for a random subset of the 350 most abundant bacteria and contrast them
with samples from the posterior predictive according to the four topic LDA model
of Section \ref{sec:antibiotics_lda} and the unigram model of Section
\ref{sec:antibiotics_unigram}. Each subpanel corresponds to a single species. The
black lines represent observed time series. Note that the $y$-axis scales vary,
as some species are much more abundant than others. Each dot is a simulated
timepoint from a posterior predictive time series. Formally, this corresponds to
the choice $T\left(x\right) = \left(x_{\cdot v}\right)_{v \in \mathcal{V}}$
where the random subset $\mathcal{V}$ indexes the displayed species.

For LDA, the posterior predictive time series are on the appropriate scale with
approximately the correct shape. However, we observe two substantial types of
departures between simulated and observed data. First, for series with larger
counts, the posterior predictive tends to oversmooth. For example, the drop to 0
in species 343 is not captured in any posterior predictive samples. Similar
oversmoothing is visible in species 1036, 1116, and 1463. A more startling type of
departure occurs in the second half of series 1116. Here, the posterior
predictive distribution places most mass on the event that the bacterial time
series rebounds to its initial abundance when in reality the species vanishes
during the second antibiotic time course, never to return. A potential
explanation for LDA's failure to capture this pattern is that few highly
abundant species disappear after the second time course, and hence they are not
captured by the global LDA summary. This suggests a technique for highlighting
outlier species: we can look at the average discrepancy between observed series
and their posterior predictive samples.

On the other hand, for the Dynamic Unigram model, the posterior predictive
distribution places most of its support close to the observed species series.
Usually, this is desirable behavior, indicating good model fit. However, here,
there is reason for concern -- the unigram model may not do much more than fit
empirical proportions at each timepoint, and there may be potential to produce
more succinct summaries that still preserve the essential structure of the data.
That is, the unigram model seems over-parameterized, simply memorizing the
input.

An alternative posterior predictive check compares PCA scores and loadings in
the true and posterior predictive data. Our motivation is that many microbiome
studies base their findings on views generated by PCA, so it would be
encouraging if our probabilistic summaries typically agree with the reductions
produced by PCA. Formally, we construct $T\left(x\right) = \left(u_{1}, u_{2},
\lambda_{1}, \lambda_{2}, v_{1}, v_{2}\right)$ where $\left(u_{i}\right)$,
$\left(\lambda_{i}\right)$ and $\left(v_{j}\right)$ are left singular vectors,
eigenvalues, and right singular vectors of $x$, respectively.

\begin{figure}[!p]
  \centering
  \includegraphics[width=\textwidth]{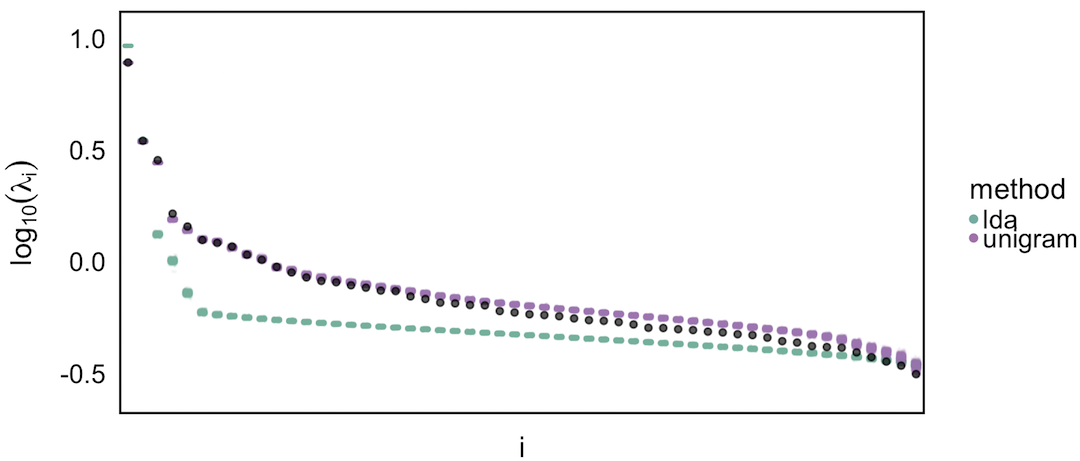}
  \caption{As a posterior predictive check, we compute eigenvalues of data
    simulated from the fitted LDA model. The clouds of points summarize the
    posterior predictive distribution, while the black circles represent
    observed data eigenvalues. Note that the $y$-axis are logged eigenvalues.
    Evidently, the four-topic model effectively creates a rank-four
    approximation of the original data. \label{fig:antibiotics_posterior_evals}}
\end{figure}

\begin{figure}[!p]
  \centering
  \includegraphics[width=\textwidth]{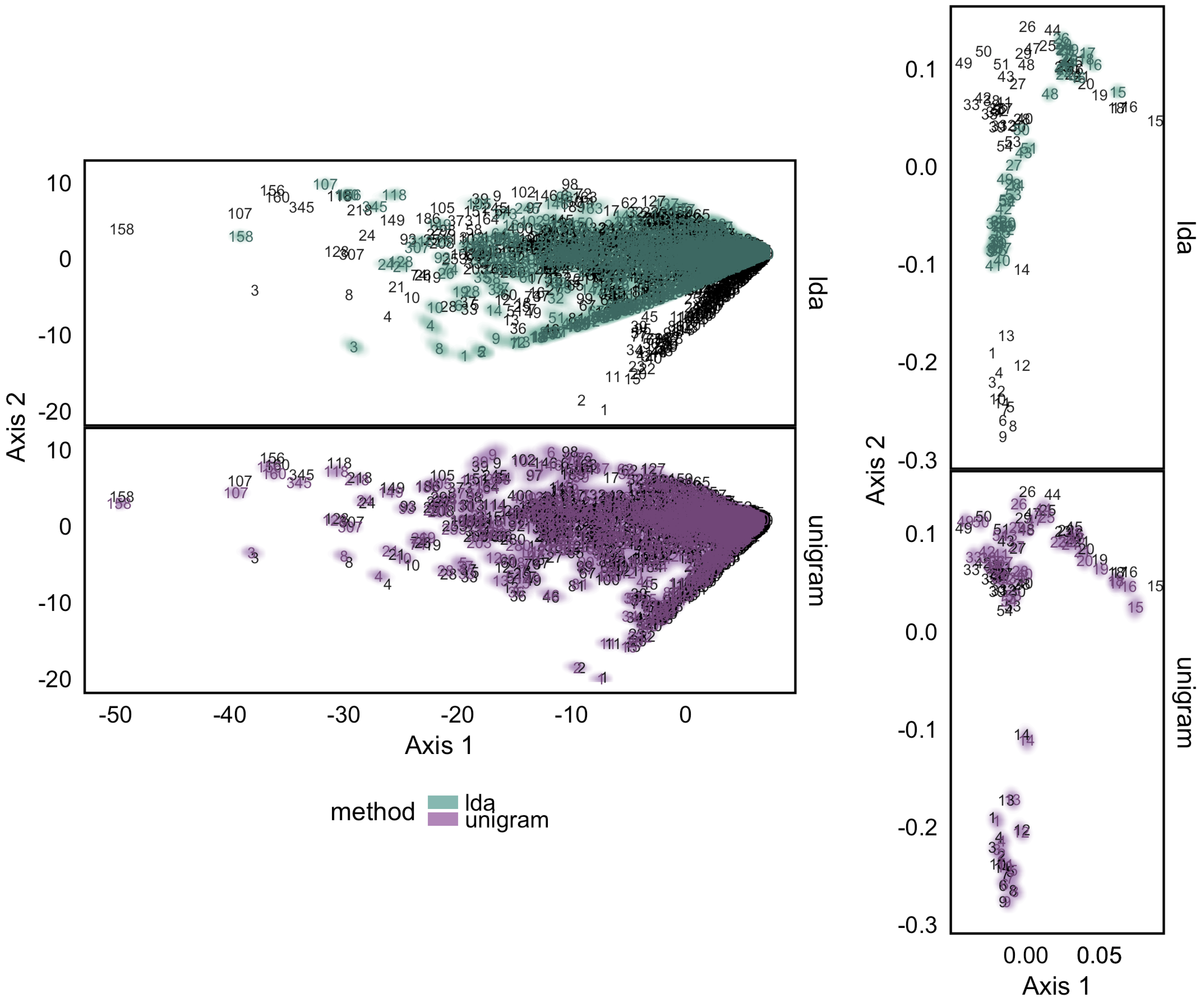}
  \caption{The eigenvalues displayed in Figure
    \ref{fig:antibiotics_posterior_evals} correspond to PCA results computed on
    posterior predictive samples, which are aligned and overlaid here. The left
    pair of panels give scores for each species, while the right pair provide
    loadings for each timepoint. The individual posterior samples have been
    smoothed into contours, while the posterior medians are displayed as shaded
    text. The observed data PCA results, after alignment with posterior samples,
    are displayed as black text. \label{fig:antibiotics_posterior_pca} }
\end{figure}

Figure \ref{fig:antibiotics_posterior_evals} gives the PCA eigenvalues between
the true and posterior predictive samples, after applying an
$\asinh$-transformation and filtering to the 1000 highest-variance bacteria. The
associated scores and loadings figures are available as Supplementary Figure
\ref{fig:antibiotics_posterior_pca}. In Figure
\ref{fig:antibiotics_posterior_evals}, each black point provides the
log-eigenvalue observed in the original data. The blue and purple clouds
consistent of small, semi-transparent, horizontally-jittered points associated
with eigenvalues from posterior predictive samples generated by LDA and the
Dynamic Unigram model, respectively. For LDA, the posterior predictive samples
have comparable top four eigenvalues, but rapidly drop-off between the fourth
and fifth eigenvalues. On the other hand, the observed data have a more steady
decline. This is likely a consequence of using $K = 4$ topics in the LDA model,
which would be consistent with the matrix factorization view of LDA described in
Section \ref{sec:nmf}. Considering these scree plots, it may be safe to increase
$K$ in follow-up analysis, as long as topics remain interpretable. In contrast,
the eigenvalues for the Dynamic Unigram model closely match those in the
observed data -- even overestimating the size of small eigenvalues -- lending
further evidence to the claim that this model is over-parameterized.

\subsection{Comparison with principal coordinates}
\label{subsec:comparison_with_pcoa}

There is value in comparing this text-modeling based approach to the data from
\citep{dethlefsen2011incomplete} with the analysis carried out in the original
publication, based on principal coordinates analysis (PCoA) using a UniFrac
distance. Both approaches accentuate the dramatic change in microbiome
composition immediately following the administration of antibiotics, and both
suggest the overall resilience of the microbiome, in the sense that the
community returns to the preantibiotic state after several days. Further, both
distinguish subject F as only making an incomplete recovery after the first
antibiotic treatment.

On the other hand, some findings are more easily accessible when using the
probabilistic approach. For example, the availability of topics with mixed
memberships strikes a balance between the continuous gradient representation of
principal coordinate analysis of \citep{dethlefsen2011incomplete} and the
discrete clusters provided by standard clustering techniques, which appear
elsewhere in the microbiome literature \citep{digiulio2015temporal,
  mcmurdie2014waste}. This simplifies the taxonomic characterization of
different communities -- with principal coordinates, it would be necessary to
find species correlated with the principal coordinate axes, and indeed no visual
representations of taxa ever appear in the figures of
\cite{dethlefsen2011incomplete}. Similarly, by studying individual topics, we
find there is more variation within taxonomic families than is ever discussed in
the original refernece. Further, through posterior predictive checks, we can
perform model assessment in a way that is not so straightforwards with PCoA, and
more broadly speaking, we by adopting probabilistic methods, we are able to
describe the uncertainty associated with mixed membership and topic estimate.

That said, PCoA enjoys certain advantages over the probabilistic approaches that
have been the focus of this work. Perhaps most importantly, PCoA can be run in
seconds, which makes it much more useful for interactive analysis. Second, by
using the UniFrac distance, PCoA is able to account for the phylogenetic
relatedness between taxa, encouraging phylogenetically similar taxa to play
similar roles in the resulting ordination. Finally, the visual representation of
samples provided by PCoA -- simply a two-dimensional scatter of the samples --
is more easily digestible than the simultaneous display of all parameters in
probability models.

Broadly speaking, it seems valuable to have both types of tools available for
practical scientific work. Ordination techniques like PCoA are useful for
describing the relationship between sets of samples, in a way that requires
little effort in either estimation or display. However, for more richly
structured summaries, which are amenable to uncertainty quantification and model
evaluation, probabilistic methods are ideal. We imagine a workflow in which
researchers quickly develop a sense of their data using ordination, and then
refine and critique their analysis using latent variables models.

\section{Discussion}

We have described the utility of taking a probabilistic modeling perspective in
the analysis of microbiome data. We have provided a detailed implementation of
benchmark analysis approaches, along with exploratory visualization of fitted
parameters and model assessment through posterior predictive checks. Through
simulation, we have established heuristics for determining the appropriateness
of applying different models and inference mechanisms, depending on the overall
data generation regime. On a real microbiome data analysis problem, we have
characterized the advantages and limitations of two probabilistic modeling
techniques. Rather than focusing on any single model, like most earlier work, we
have emphasized the practice of contrasting, critiquing, and learning from
multiple alternatives. Throughout, we have emphasized both insights, in terms of
estimated parameters, as well as uncertainty, in the form of full approximate
posterior distributions. We hope our efforts help to widen the biostatistician's
toolbox and clarify the practical advantages and limitations of different
approaches to microbiome data.

Microbiome studies are a source of richly structured, high-dimensional data,
coupled with novel scientific problem setups. For example, in the antibiotics
data set described here, we have already encountered structure in the form of
zero-inflated counts, time series with changepoints, and apriori known
phylogenetic relationships between features. It is anticipated that future
microbiome studies will collect an increasing number of samples as well as more
data sources per sample -- spectral and genomic, in addition bacterial
abundance, for example \citep{jansson2016multi}. Further, the investigations
often revolve around a combination of ecological community characterization and
medically-relevant identification of treatment effects. We believe we have only
begun to see the potential for probabilistic methods to guide careful scientific
reasoning -- which emphasizes both insights and the degree of uncertainty about
them -- in these complex scenarios.

\section{Supplementary Material}

Supplementary material is available online at
\url{http://biostatistics.oxfordjournals.org}.

\section{Reproducibility}

Code for all simulations, data analysis, and figures is available at
\url{https://github.com/microbiome\_plvm}. Detailed instructions are available
in the repository README.md. Further, a docker image with all software
requirements preinstalled is available from
\url{https://hub.docker.com/r/krisrs1128/microbiome_plvm/}, and the
corresponding Dockerfile is provided in the github repository.

\section*{Acknowledgments}

KS is supported by a Stanford University Weiland fellowship and the National
Institute of Health T32 grant 5T32GM096982-04. SPH is supported by the National
Institute of Health TR01 grant AI112401.

{\it Conflict of Interest}: None declared.

\bibliographystyle{biorefs}
\bibliography{refs}

\section{Supplementary Figures}

\begin{figure}[ht]
  \centering
  \includegraphics[width=\textwidth]{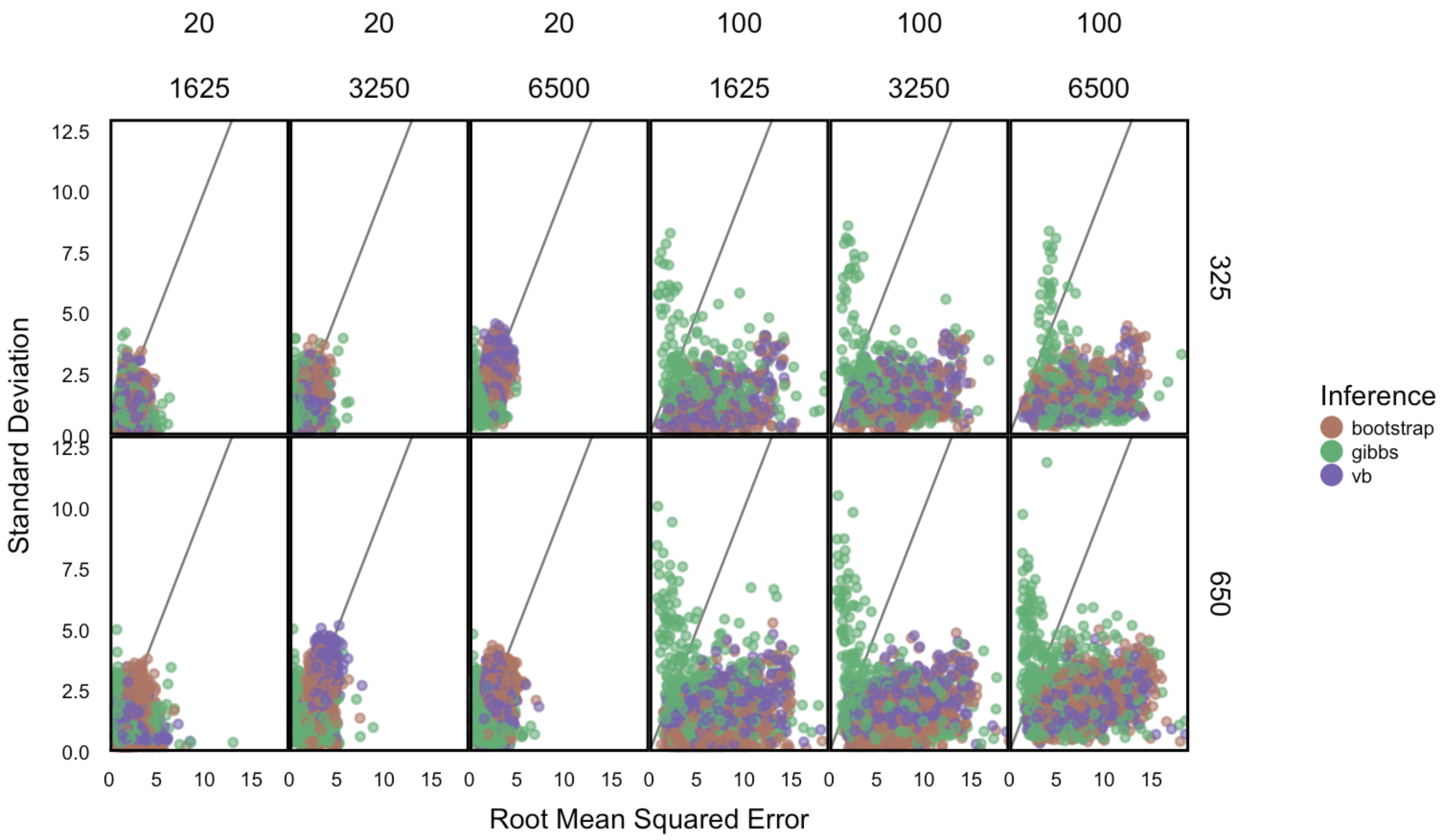}
  \caption{A simplification of Figure \ref{fig:mu_intervals}, displaying RMSE when
    using posterior medians to estimate simulation $\mu_{tv}$s ($x$-axis) and
    the standard deviations of posterior marginals ($y$-axis), across
    experimental configurations. Generally, MCMC sampled posteriors seem to be
    the most reliable, across simulation configurations.
    \label{fig:mu_errors_unigram}
  }
\end{figure}

\begin{figure}[!p]
  \centering
  \includegraphics[scale=0.2]{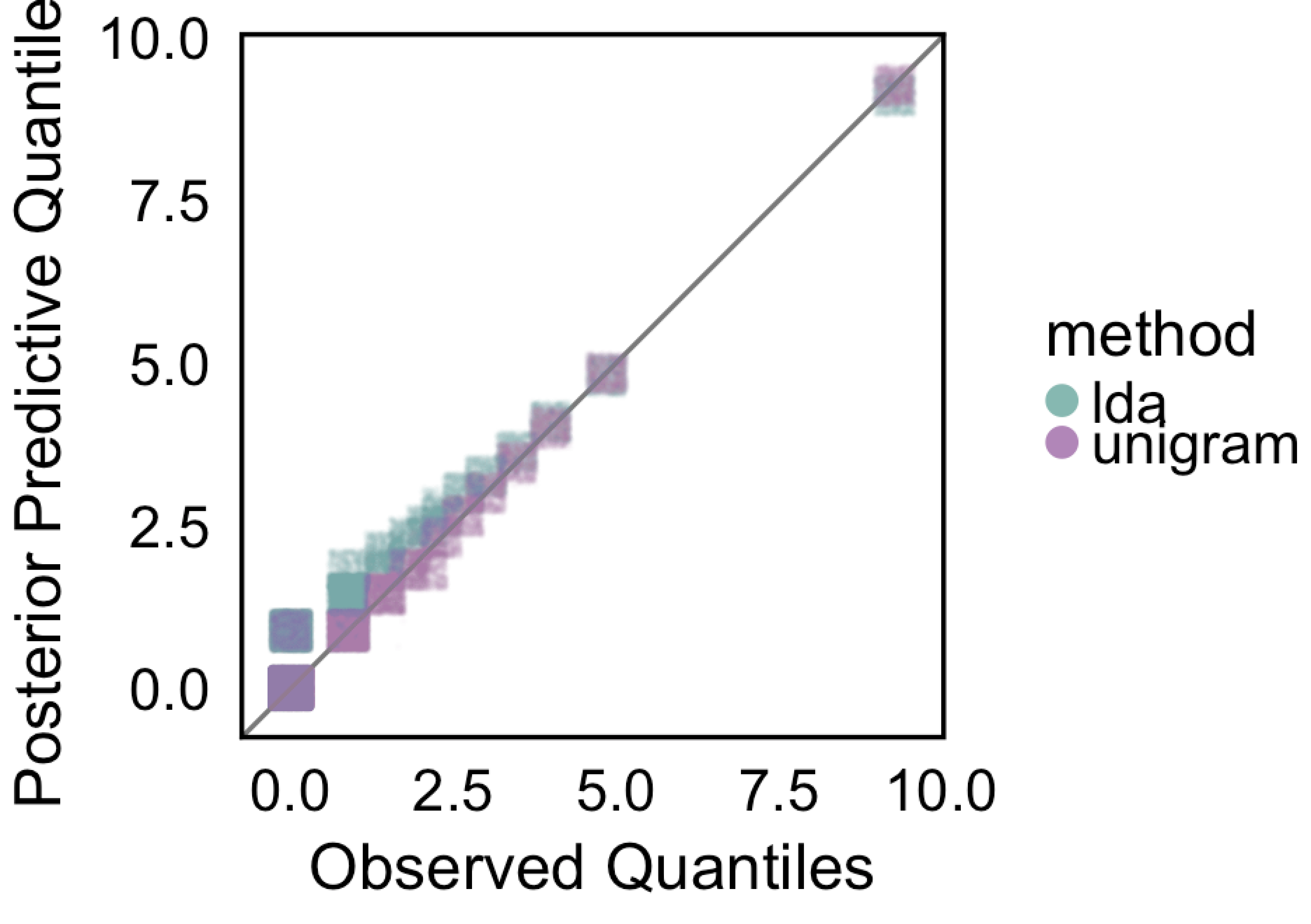}
  \caption{As a posterior check, we compare the observed with simulated
    data quantiles, using a qq-plot. To reduce overlap, we have introduced a
    uniform $\left[0, 0.2\right]$ jitter on both axes. Further, the points are
    semi-transparent -- this makes it easy to see that most quantiles map to 0,
    which is expected, considering the sparsity of the data. From this view, we
    see that the LDA model tends to underestimate the overall number of zeros in
    the data, while the Dynamic Unigram model matches the observed quantiles
    almost exactly. \label{fig:antibiotics_posterior_quantiles} }
\end{figure}

\begin{figure}[!p]
  \centering\includegraphics[width=\textwidth]{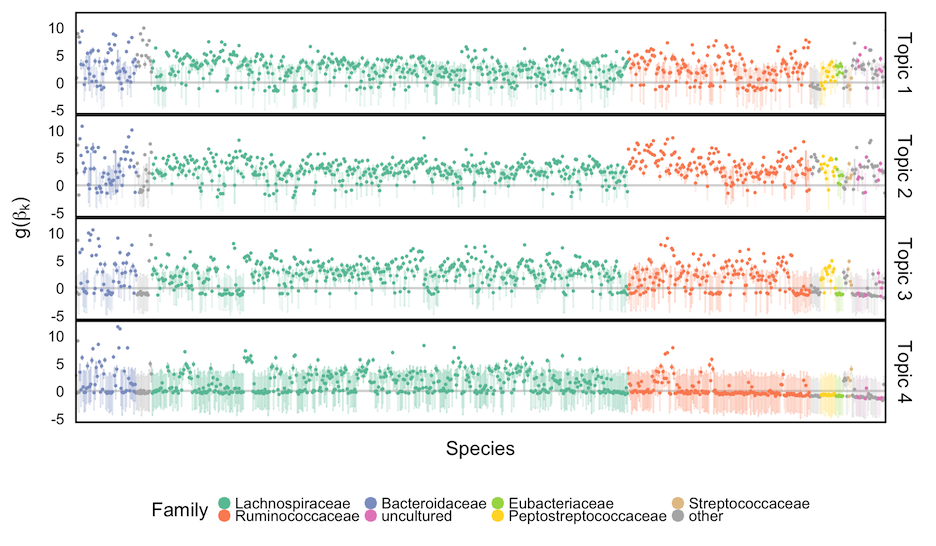}
  \caption{The analog of Figure \ref{fig:antibiotics_lda_beta} for Subject D.}
\end{figure}

\begin{figure}[!p]
  \centering\includegraphics[width=\textwidth]{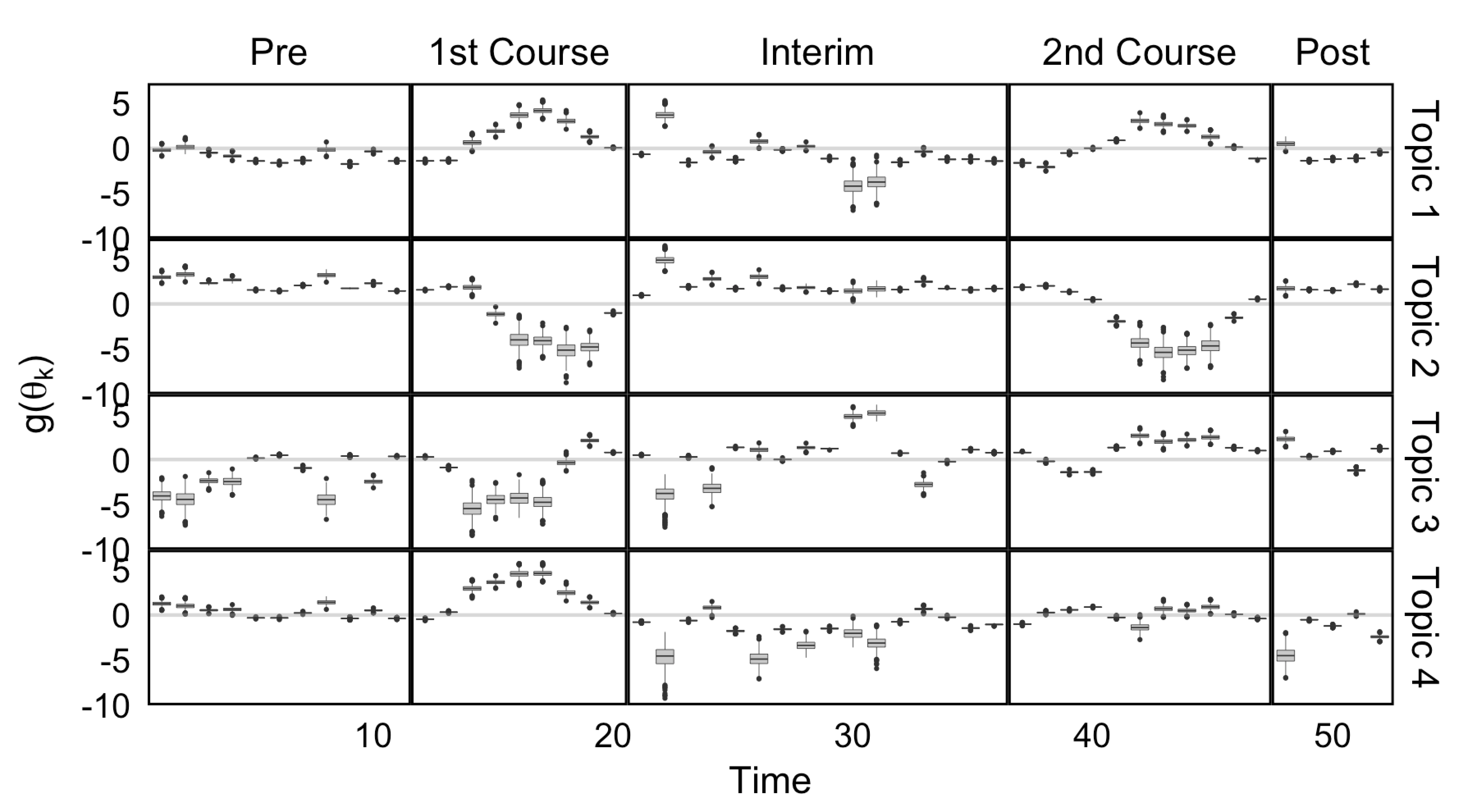}
  \caption{The analog of Figure \ref{fig:antibiotics_lda_theta} for Subject E.}
\end{figure}

\begin{figure}[!p]
  \centering
  \includegraphics[width=0.8\textwidth]{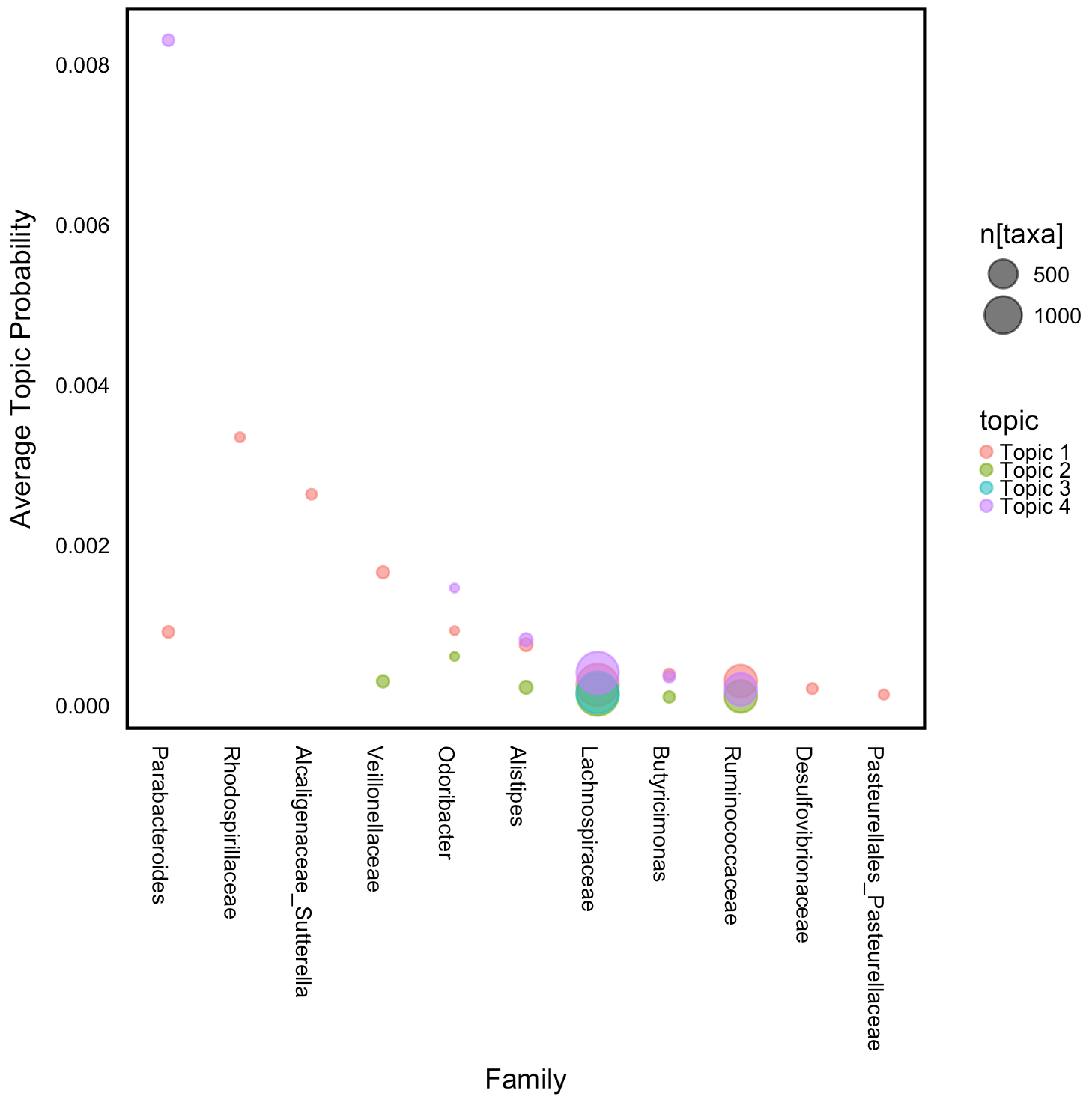}
  \caption{We can search for entire taxonomic families that seem associated with
    individual topics. Here we calculate the average of the topic
    representativeness statistic $\beta_{kv} - \sum_{k^{\prime} \neq k}
    \beta_{k^{\prime} v}$ across all species $v$ within each
    Family. Only those families that are most associated with a topic are
    displayed here. The sizes of circles represents the number of species within
    the family, which can be used to gauge the variability of the
    estimate. Compare this view with Supplementary Figure
    \ref{fig:uneven_taxa_facet}. \label{fig:uneven_taxa_ordered} }
\end{figure}

\begin{figure}[!p]
  \centering
  \includegraphics[width=\textwidth]{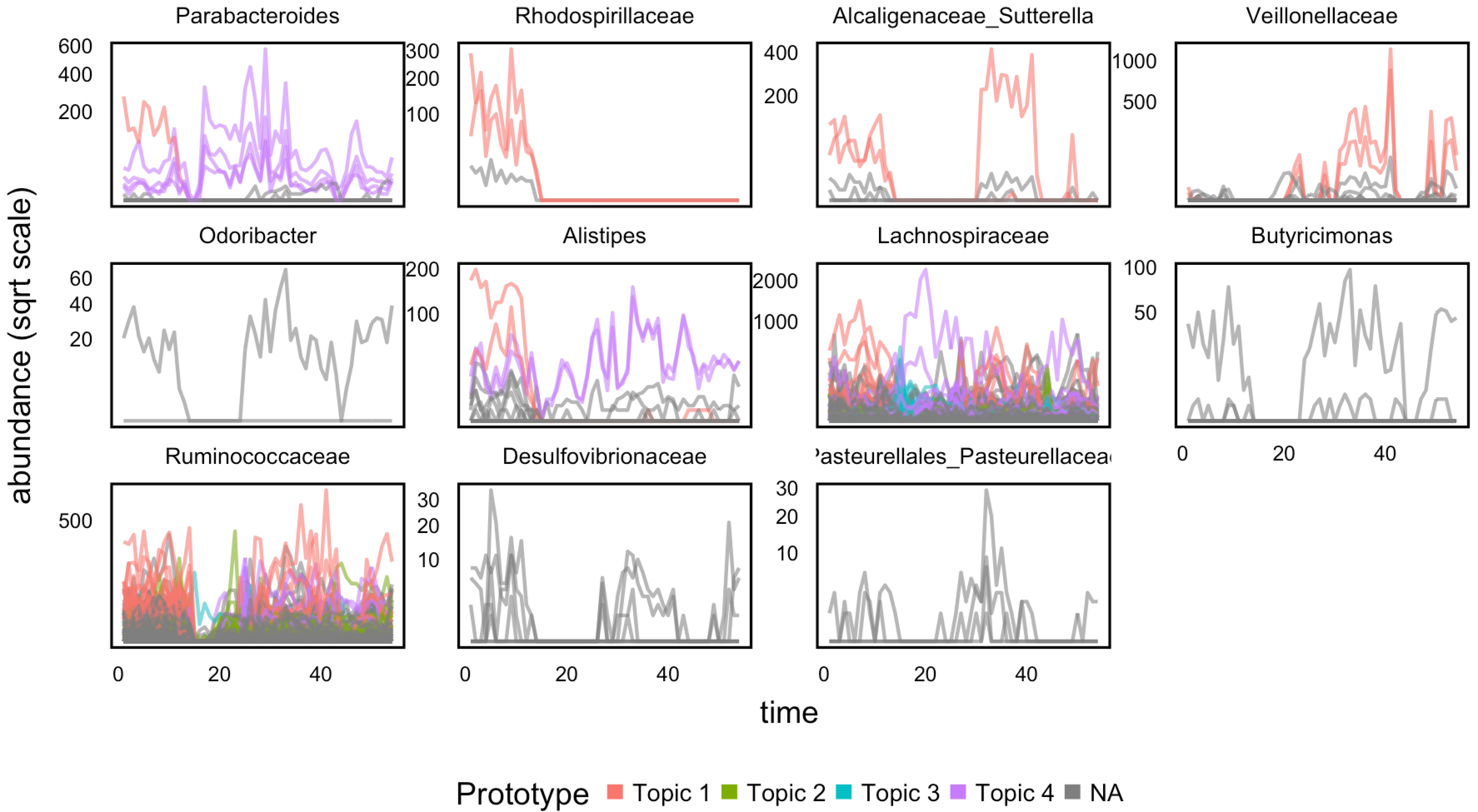}
  \caption{All species within the families screened out from Figure
    \ref{fig:uneven_taxa_ordered} are displayed here. Species among the
    representatives displayed in Figure \ref{fig:topic_prototypes} are colored
    according to the topics of they are prototypical. Grey series still
    contribute to the average family-topic association measure, but were not
    among the 50 prototypes for each topic. This view suggests that the
    Rhodospillaceae, Alcaligenaceae and Parabacteroides may have a large
    fractions of representatives from Topics 3, 1, and 2, respectively. Note
    that as before, abundances are plotted on a square root
    scale. \label{fig:uneven_taxa_facet} }
\end{figure}
\end{document}